\documentclass[oldversion]{aa}
\usepackage{times}
\usepackage{mathptmx}
\usepackage{multirow}
\usepackage{natbib}
\usepackage{amssymb,amsfonts}
\usepackage{rotating}
\usepackage[]{fontenc}
\usepackage[latin1]{inputenc}
\usepackage[]{graphicx}

\begin{document}

\title{Energetics of the molecular gas in the H$_2$ luminous radio galaxy 3C~326: 
Evidence for negative AGN feedback
\thanks{Based on observations carried out with the IRAM Plateau de
 Bure Interferometer}}

\author{N.~P.~H.~Nesvadba,\inst{1,2}, F.~Boulanger\inst{1},
  P.~Salom\'e\inst{3,4}, P. Guillard\inst{1}, M. D. Lehnert\inst{5}, P. Ogle\inst{6},
P. Appleton\inst{7}, E. Falgarone\inst{4,8} \and G. Pineau des Forets\inst{1,4}}    

\institute{Institut d'Astrophysique Spatiale, CNRS, Universit\'e Paris Sud,
91405 Orsay, France 
\and
email: nicole.nesvadba@ias.u-psud.fr
\and
Institut de Radioastronomie Millim\'etrique (IRAM), St. Martin d'Heres, France
\and
LERMA, Observatoire de Paris, CNRS, Paris, France
\and
GEPI, Observatoire de Paris, CNRS, Universit\'e Denis Diderot, Meudon, France
\and
Spitzer Science Center, California Institute of Technology, Pasadena, USA
\and 
NASA Herschel Science Center, California Institute of Technology, Pasadena, USA
\and
\'Ecole Normale Sup\'erieure and Observatoire de Paris, Paris, France
}

\authorrunning{Nesvadba et al.}
\titlerunning{The multiphase gas in a H$_2$ luminous radio galaxy}

\date{Received  / Accepted }

\abstract
{
We present a detailed analysis of the gas conditions in the H$_2$
luminous radio galaxy 3C~326~N at z$\sim$0.1, which has a low star-formation rate (SFR$\sim$0.07 M$_{\odot}$ yr$^{-1}$) in spite of a gas
surface density similar to those in starburst galaxies. Its
star-formation efficiency is likely a factor $\sim$10-50 lower than
those of ordinary star-forming galaxies.  Combining new IRAM CO
emission-line interferometry with existing Spitzer mid-infrared
spectroscopy, we find that the luminosity ratio of CO and pure
rotational H$_2$ line emission is factors 10-100 lower than what is
usually found. This may suggest that most of the molecular gas is
warm. The Na~D absorption-line profile of 3C~326~N in the optical
suggests an outflow with a terminal velocity of $\sim -$1800 km
s$^{-1}$ and a mass outflow rate of 30-40 M$_{\odot}$ yr$^{-1}$, which
cannot be explained by star formation. The mechanical power implied by
the wind, of order $10^{43}$ erg s$^{-1}$, is comparable to the
bolometric luminosity of the emission lines of ionized and molecular
gas. To explain these observations, we propose a scenario where a
small fraction of the mechanical energy of the radio jet is deposited
in the interstellar medium of 3C~326~N, which powers the outflow,
and the line emission through a mass, momentum and energy exchange
between the different gas phases of the ISM. Dissipation times are of
order $10^{7-8}$ yrs, similar or greater than the typical jet
lifetime. Small ratios of CO and PAH surface brightnesses in another 7
H$_2$ luminous radio galaxies suggest that a similar form of AGN
feedback could be lowering star-formation efficiencies in these
galaxies in a similar way. The local demographics of radio-loud AGN
suggests that secular gas cooling in massive early-type galaxies of
$\ge 10^{11}$ M$_{\odot}$ could generally be regulated through a fundamentally
similar form of 'maintenance-phase' AGN feedback.
}


\maketitle

\section{Introduction} 
\label{sec:introduction}

Molecular gas plays a critical role for our growing understanding of galaxy
evolution. It often dominates the mass budget of the
interstellar medium in galaxies, and is most closely related to the
intensity at which galaxies form stars \citep[e.g.,][]{kennicutt98}. Being
strongly dissipative, it is also particularly susceptible to
the astrophysical processes that drive galaxy evolution -- interactions, or
feedback from starbursts and AGN -- and therefore plays a key role for our
understanding of how these processes regulate star formation and galaxy
assembly.

It has only recently been recognized that powerful AGN may play a significant
role in regulating galaxy growth over cosmological timescales by suppressing
gas accretion and star formation \citep[e.g.,][]{silk98,friaca98,scannapieco04,springel05,bower06,croton06,ciotti07,merloni08}. Such AGN 'feedback' would help resolve
some of the remaining discrepancies between hierarchical models of galaxy
evolution -- implying a rather gradual assembly of massive galaxies -- and
observations, which suggest that massive galaxies formed most of their stars
at high redshift, whereas star formation at later epochs was strongly
suppressed. Observationally, a picture is emerging where radio jets may play a
large role in transforming the energy ejected by the AGN into kinetic and
thermal energy of the interstellar medium of the host galaxy. Observations of
radio-loud AGN \citep[e.g.,][]{heckman91, heckman91b, morganti05, emonts05,
best05, best06, nesvadba06, nesvadba07, mcnamara07, nesvadba08, holt08,
baldi08, fu09, humphrey09} and a large number of hydrodynamical simulations
\citep[e.g.,][]{krause05, saxton05, heinz06, sutherland07, merloni07,
antonuccio08} suggest that radio-loud AGN inject a few percent of their
mechanical energy into the ambient gas, parts of which produce significant
outflows of warm gas \citep{morganti03, morganti05, emonts05, nesvadba06,
 nesvadba07, holt08, nesvadba08, fu09}. However, most previous studies
focused on the warm and hot gas at temperatures $\ge 10^4$ K, and did not
address the impact on the molecular phase, which is a serious limitation if we
want to understand how the radio-loud AGN may regulate star formation in the
host galaxy.

Observations with the Spitzer IRS spectrograph recently revealed a
significant number of ``H$_2$-luminous'' galaxies, where the molecular
gas does not appear associated with star formation \citep[][see also
  \citealt{haas05}]{appleton06, egami06, ogle07, ogle08, ogle09,
  demessieres09, sivandam09}\footnote{\citet{ogle07, ogle09} propose
  to introduce a new empirical classification based on this H$_2$
  excess, and refer to such targets as ``Molecular Hydrogen Emission
  Galaxies'', MOHEGs}.  The mid-infrared spectra of H$_2$-luminous
galaxies are dominated by bright, pure rotational emission lines of
warm molecular hydrogen, (${\cal L}(H_{2}) =10^{40}$ -- $10^{43}\,
{\rm erg \, s^{-1}}$), while classical star-formation indicators like
a bright infrared continuum, mid-infrared lines of [NeII] and [NeIII],
and PAH bands are weak or absent.  Interestingly, \citet{ogle09} find
that 30\% of their radio-loud AGN taken from the 3CR are H$_2$
luminous, suggesting this may be a common phenomenon which could be
related to interactions with the radio source.

To test this hypothesis and to evaluate possible consequences for the
gas properties and star formation in the host galaxy, we have started
CO emission-line observations of H$_2$ luminous radio galaxies with
the IRAM Plateau de Bure Interferometer. Our goal is to constrain the
physical properties and masses of the multiphase warm and cold gas in
these galaxies, and to measure the gas kinematics. Here we present a
detailed analysis of the multiphase gas content, energetics, and
dissipation times of the H$_2$-luminous radio galaxy 3C~326~N at
z=0.09 \citep{ogle07,ogle08,ogle09}, which has particularly high
H$_2$/PAH ratios. This analysis is based on our new CO(1-0) observations,
as well as existing mid-infrared Spitzer and SDSS optical
spectroscopy. Specifically we address three questions: What powers the
H$_2$ emission in this galaxy? What is the physical state of the
molecular gas, and perhaps most importantly, why is 3C~326~N not
forming stars?

We find that the interstellar medium of 3C~326~N has very unusual
physical properties, where the warm molecular gas may dominate the
overall molecular gas budget (\S\ref{ssec:massbudget}) and where the
emission-line diagnostics suggest that the molecular as well as the
ionized gas may be mainly excited by shocks (\S\ref{sec:diagnostics})
giving rise to luminous line emission at UV to mid-infrared
wavelengths. We also identify a significant outflow of neutral gas
from Na~D absorption profiles, which cannot be explained by star
formation (\S\ref{sec:outflow}). We propose a physical framework in
which these observations can be understood as a natural consequence of
the energy and momentum coupling between the gas phases, which is
driven by the mechanical energy injection of the radio jet
(\S\ref{sec:enmom}). This scenario is an extension of the classical
'cocoon' model \citep[e.g.,][]{scheuer74,begelman89} explicitly taking
into account the multiphase character of the gas, with an emphasis on
the molecular gas. We use our observational results to quantify some
of the parameters of this scenario, including, perhaps most
importantly, the dissipation time of the turbulent kinetic energy of
the gas, and find that it is self-consistent and in agreement with the
general characteristics of radio-loud AGN.

3C~326~N shows evidence for a low star-formation efficiency leading to
a significant offset from the Schmidt-Kennicutt relationship of
ordinary star-forming galaxies by factors 10$-$50, which is similar to
other H$_2$ luminous radio galaxies with CO observations in the
literature, as would be expected if 3C~326~N was a particularly
clear-cut example of a common, underlying physical mechanism that is
throttling star formation (\S\ref{sec:SF}). Long dissipation times
suggest that the gas may remain turbulent over timescales of
$10^{7-8}$ years, of order of the lifetime of the radio source, or
perhaps even longer (\S\ref{ssec:dissipationtime}), while the energy
supplied by the radio source may be sufficient to keep much of the gas
warm for a Hubble time (\S\ref{sec:implications}) as required during
the maintenance phase of AGN feedback, assuming typical duty cycles of order $10^8$ yrs.

Throughout the paper we adopt a H$_0 =$70 km s$^{-1}$, $\Omega_{M}=0.3$,
$\Omega_{\Lambda} =$0.7 cosmology. In this cosmology, the luminosity distance
to 3C~326~N is D$_L^{326N}=$ 411 Mpc. One arcsecond corresponds to a
projected distance of 1.6 kpc.

\section{Observations} 
\label{sec:observations}
\subsection{The H$_2$ luminous radio galaxy 3C~326~N}
\label{ssec:targets}
We present an analysis of the powerful H$_2$ luminous radio galaxy
3C~326~N at z$\sim$0.1 \citep{ogle07,ogle09}.  Pure rotational,
mid-infrared H$_2$ lines in 3C~326N have an extraordinary luminosity
and equivalent width \citep{ogle07}. The total emission-line
luminosity of 3C~326N is $\rm L(H2) = 8.0 \pm 0.4 \times 10^{41} \,
erg\, s^{-1}$ (integrated over the lines S(1) to S(7)), corresponding
to $17 \pm 2$\% of the infrared luminosity integrated from 8-$70\ \mu
$m. This is the most extreme ratio found with Spitzer so far. The
H$_2$ line emission is not spatially or spectrally resolved, implying
a size $<$4\arcsec\ ($\sim$6 kpc at the distance of the source), and a
line width FWHM$\le$2500 km s$^{-1}$.

3C~326~N is remarkable in that it does not show evidence for strong star
formation, in spite of a significant molecular gas mass of $\sim 10^9$
M$_{\odot}$ \citep{ogle07,ogle09}. This mass corresponds to a mean surface
density of warm H$_2$ within the slit of the short-wavelength spectrograph of
Spitzer of $\rm 30 \, M_\odot \, pc^{-2}$, a few times larger than the total
molecular gas surface density of the molecular ring of the Milky Way
\citep{bronfman88}. Generally, galaxies with similarly large molecular gas
masses show vigorous starburst or AGN activity. However, in 3C~326~N the
luminosity of the PAH bands and 24 $\mu$m dust continuum suggest a
star-formation rate of only $\sim 0.07$ M$_{\odot}$ yr$^{-1}$, about 2\% of
that in the Milky Way. Given this large amount of molecular gas and the high
mass surface density, the bolometric AGN luminosity of 3C~326~N is also
remarkably low.

3C~326~N has a Mpc-sized FRII radio source, which is relatively weak
for this class, with a radio power of $2.5\times 10^{26}$ W Hz$^{-1}$
at 327 MHz. There is some confusion in the literature regarding which
  galaxy, 3C~326~N or the nearby 3C~326~S is associated with the radio
  lobes. Both candidates have detected radio cores and the extended
  radio lobes make it difficult to uniquely associate one or the other
  component with the radio source. \citet{rawlings90} argue that
  3C~326~N is the more plausible candidate, having the brighter stellar
  continuum \citep[and greater stellar mass,][]{ogle07}. Only 3C~326~N
  has luminous [OIII]$\lambda$5007 line emission consistent with the
  overall relationship between [OIII]$\lambda$5007 luminosity and
  radio power \citep{rawlings90}. The core of 3C~326~N appears unresolved at 8.5
  GHz with a 2\arcsec\ beam \citep{rawlings90}. We discuss
  the age and kinetic power of the radio source of 3C~326
  \S\ref{ssec:radiojet}.

\subsection{CO millimeter interferometry}
\label{ssec:co}

CO(1--0) emission-line observations of 3C~326 were carried out with the IRAM
Plateau de Bure Interferometer (PdBI) in two runs in January and July/August
2008 with different configurations. In January, we used the narrow-band and
dual-polarisation mode, corresponding to a band width of 950 MHz, or 2740 km
s$^{-1}$, with a channel spacing of 2.5 MHz. The 6 antennae of the PdBI were
in the BC configuration at a central frequency of 105.85 GHz, corresponding to
the observed wavelength of the CO(1--0) line at a redshift of z=0.090.  The
final on-source integration time for this run was 6.7 hrs (discarding scans
with atmospheric phase instabilities). The FWHP of the synthesized beam in the
restored map is 2.5$^{\prime\prime}$ $\times$ 2.1$^{\prime\prime}$ with a
position angle of 23$^\circ$.  Only data from this run were used for the CO
emission-line measurements.

The presence of a 3~mm non-thermal continuum from the radio source and
relatively narrow receiver bandwidth made it difficult to rule out a possible
contribution of a very broad CO(1--0) line (FWHM$>$1000 km s$^{-1}$) from the
January 2008 data alone. We therefore re-observed 3C~326~N in July and August
2008 with the goal of carefully estimating the continuum level, taking
advantage of the 1.75 GHz bandwidth in the wide-band single-polarisation mode,
corresponding to 4956 km s$^{-1}$. For these observations we used 5 antennae
in the compact D-configuration. The final on-source integration time for this
run was 5.2 hours (again, discarding scans with strong atmospheric phase
instabilities). We detected the millimeter continuum at a level of
S$_{3mm}$=0.99$\pm$0.05 mJy at the centimeter position of the radio
source measured by \citet{rawlings90}. 

Data reduction and analysis relied on GILDAS \citep{pety05}. The flux was
calibrated against MWC349 and against reference quasars whose flux is
monitored with the PdBI. We removed the continuum by assuming a point source
at the position of 3C~326~N with a flat spectrum of 1 mJy in the uv-plane.
The data were then averaged in 71 km s$^{-1}$-wide velocity channels. 

\subsection{CO line emission}
\label{ssec:coin326}
\begin{figure}
\centering \includegraphics[width=0.5\textwidth]{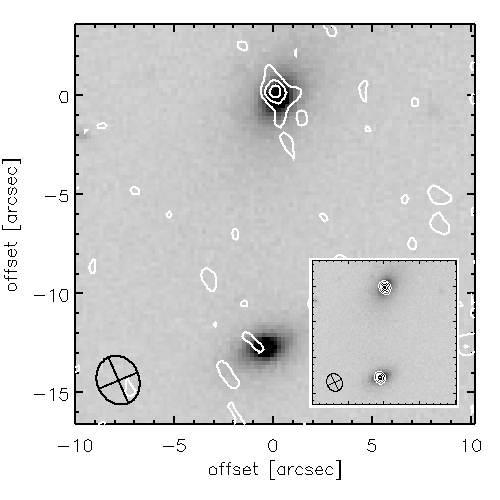}\\
\caption{Continuum-free CO(1-0) emission-line morphology of 3C~326~N (white
  contours) superimposed on the SDSS R-band image of 3C~326~N and
  3C~326~S. Contours are given for 3, 4, and 5$\sigma$. The inset
  shows the same SDSS R-band with the line-free 3~mm continuum morphology
  superimposed as white contours. We used the continuum morphologies to align
  the millimeter and optical data, assuming that the AGN reside in the nuclei
  of the galaxies. With this alignment, we find an offset of
  0.7\arcsec\ between the CO line emission and continuum (millimeter and
  optical) which, given the faintness of the line emission and much greater
  beam size, is not significant. The ellipse in the lower left corner shows the
  beam size and position angle in both images.} 
\label{fig:CO_map}
\end{figure}

We show the continuum-free CO(1--0) emission-line map of 3C~326~N in the inset
of Figure~\ref{fig:CO_map}, integrated over 637 km s$^{-1}$ around the
line centroid.  The central position is slightly offset from the millimeter
continuum position (by 0.37\arcsec$\times$0.80\arcsec\ in right ascension and
declination, respectively), corresponding to 1/3 to 1/6 of the beam size,
respectively. Given the relatively low signal-to-noise ratio, this offset is
not significant.

We measured the CO(1--0) line flux in different apertures and in the uv
plane to investigate 
whether the line emission may be spatially extended, finding a larger flux for
elliptical Gaussian models than for a point source.  This may suggest that the
source is marginally spatially extended with a size comparable to our
2.5\arcsec$\times$2.1\arcsec\ beam. Results for the spectrum integrated over
different apertures are summarized in Tables \ref{tab:co1} and \ref{tab:co2}.

\begin{figure}
\centering
\includegraphics[width=0.8\textwidth,angle=0]{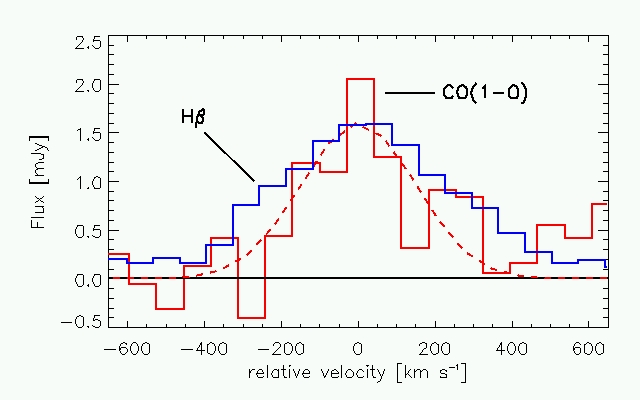}\\
\caption{CO(1-0) millimeter spectrum of 3C~326N (red solid line), and H$\beta$
line profile (blue solid line). The red dashed line shows our best fit to
the CO(1-0) line profile. Both spectra are continuum-subtracted.}
\label{fig:CO_spectrum}
\end{figure}

In Figure~\ref{fig:CO_spectrum} we show our CO(1--0) spectrum of
3C~326~N integrated over a 5\arcsec\ aperture. The blue line shows the
Gaussian line fit which gives a FWHM=$350\pm$100 km s$^{-1}$ and an integrated 
emission-line flux of I$_{CO}$=1.0$\pm$0.2 Jy km s$^{-1}$ (over a
5\arcsec\ aperture, corresponding to the slit width of the IRS spectrum). The
observed frequency corresponds to a redshift of z$=$0.0901$\pm$0.0001 and is
similar to the observed redshift of the optical absorption lines within the
uncertainties.

\subsection{Optical spectroscopy}
\label{ssec:obsoptspec}
We complement these new millimeter observations of the cold molecular gas with
publicly available optical spectroscopy from the Sloan Digital Sky Survey
\citep{york00}. This spectrum traces the warm ionized gas and the stellar
continuum. It has signal-to-noise ratios of $\sim$20-30 in the continuum
(measured at wavelengths around 5000-6000 \AA). 3C~326~N is significantly
larger than the 3\arcsec\ diameter of the SDSS fibers, hence these are
near-nuclear spectra including light from the central 5 kpc. This is well
matched to the 2.5\arcsec$\times$2.1\arcsec\ beam size of our IRAM millimeter
observations, and to the 5\arcsec\ slit width of the Spitzer-IRS spectra.  The
data cover the spectral range between [OII]$\lambda\lambda$3726,3729 and
[SII]$\lambda\lambda$6716,6731 emission lines and the Ca~H$+$K, Mg~b, Na~D
absorption lines. All lines are relatively broad and spectrally well resolved
at the spectral resolving power of the SDSS of R$\sim$1800 \citep{york00}. All
measured line properties are summarized in Table~\ref{tab:3c326emlines}. While
the low resolution makes it difficult to resolve certain line doublets, it
does allow us to measure velocity dispersions and to analyze the line
profiles.

We used the publicly available \emph{Starlight} package \citep{cid05} to fit
the stellar continuum of 3C~326~N, and to correct our emission-line
measurements for the underlying continuum emission and stellar absorption
lines. \emph{Starlight} allows to fit the optical stellar continuum of a
galaxy with a linear combination of simple stellar populations (SSPs) of
different ages and metallicities. We used SSPs from the stellar population
synthesis models of \citet{bc03} for a wide range of ages (from $\sim 10^{7}$
yrs to several $10^{10}$ yrs) and metallicities (Z$=$0.0002-0.05).
Wavelengths near strong emission lines are excluded from the fit.
Formally, we give our best-fit parameters. We find that 97\% of the
population formed 13 Gyr ago, with the remaining 3\% formed 1.3 Gyr
ago. However, visual inspection of a fit with or without the
intermediate-age population reveals no difference. Only populations
with solar or supersolar metallicity contribute. Adopting a Calzetti
extinction law we find A$_V =$0.17 mag. In light of the unavoidable
degeneracies and systematic uncertainties, we caution that these fits
may be somewhat less constraining than the numerical results suggest
at face value. This does not change the basic result, that overall,
these fits do imply an old stellar population with age $>$10 Gyr, solar
metallicity, and low extinction. We show the spectrum of 3C~326~N
along with the best fit for the most age-sensitive part around the
4000\AA\ spectral break in Fig.~\ref{fig:stpops}.  Unless stated
otherwise, we will in the following only refer to the
continuum-subtracted emission lines.

\subsection{Emission-line kinematics}
\label{ssec:emlinekin}

After subtracting the underlying continuum emission, we fitted the emission
lines with Gaussian profiles, where line widths, fluxes, and redshifts
are free parameters, except for the [OII]$\lambda\lambda$3726,3729;
[OIII]$\lambda\lambda$4959,5007, [NII]$\lambda\lambda$6548,6583, and
[SII]$\lambda\lambda$6716,6731 doublets, where we required the redshifts and
line widths in the two lines of each doublet to be identical. For
[OIII]$\lambda\lambda$4959,5007, [NII]$\lambda\lambda$6548,6583 we also
required a flux ratio of 1/3 between the fainter and the brighter component.
Results for individual lines are given in Table~\ref{tab:3c326emlines}.

All lines have relatively large widths with FWHM$\sim$600 km s$^{-1}$. Redshifts and
line widths are similar within a scatter of $\sim 30$ km s$^{-1}$ for all
lines. Redshifts are consistent with the systemic redshift of
z$=$0.0900$\pm$0.0001 which we obtained from the stellar absorption lines.

Careful inspection of the profiles of the relatively luminous, and
spectrally well isolated [OIII]$\lambda\lambda$4959,5007 and
[OI]$\lambda\lambda$6300,6363 lines reveals the presence of very broad
components with widths of FWHM$=$1635 km s$^{-1}$
(Fig.~\ref{fig:optlinefits}). Broad components of forbidden lines
are not associated with typical nuclear broad-line regions
\citep{sulentic00} but do suggest that some of the ISM in 3C~326~N
is kinematically strongly disturbed. The redshifts and line width of
the very broad components are overall consistent with the wind component of the
Na D line discussed in \S\ref{sec:outflow}. Equally large line
widths of ionized gas have also been found by \citet{holt08} in
nearby radio galaxies showing the signs of AGN feedback and by
\citet{nesvadba06,nesvadba07,nesvadba08} in the kpc-scaled jet-driven outflows
of ionized gas in high-redshift radio galaxies.

To compensate for the difficulty of fitting broad components at relatively low
signal-to-noise ratios with more narrow, superimposed components, we required
that all four lines have the same redshift and line widths, which yields
reasonably good residuals. For the broad lines, we do not find a significant
blueshift or redshift relative to the narrow components.

Due to the large line widths, several lines are blended, which makes a
detailed fit more difficult (namely, these are [OII]$\lambda3726,3729$,
H$\alpha$ and [NII]$\lambda\lambda$6548,6583, and [SII]$\lambda
\lambda$6716,6731). We did not attempt to fit multiple components to these
lines.

\subsection{Characteristics of the Na D lines}
\label{ssec:NaD}
Figure~\ref{fig:3C326NaDspec} shows the strong and broad Na~D absorption
feature that we detect in our SDSS spectrum of 3C~326~N.  We used the
stellar population synthesis models described in \S\ref{ssec:obsoptspec} 
to estimate and remove the stellar component of the Na~D line which
can be significant in stellar populations. We also show Mg~b in
Figure~\ref{fig:3C326NaDspec}, which is purely stellar and illustrates
the accuracy of the population synthesis fit.

Figure~\ref{fig:3C326NaDspec} shows that the Na~D lines in 3C~326~N
are very broad, so that we cannot directly measure the equivalent
width of each component of the Na D doublet. We therefore modeled the
lines using the atomic data compiled in \citet{morton91}, allowing for
changes in the turbulent velocity, velocity offsets, and covering
fraction. Since we do not resolve individually the lines of the
doublet, covering fraction (C$_f$) and optical depth are degenerate
unless the line shapes are accurately known.  The lines are heavily
blended and of insufficient S/N to have confidence in the line shapes.

We find that the Na~D lines are heavily saturated so that we can only
give lower limits on the Na I column densities. Values in the range 4$-$10
$\times$10$^{14}$ cm$^{-2}$ (for C$_f$=1 and 0.5 respectively) are most
likely. The Doppler parameter of the doublet is $\sim$800 km s$^{-1}$
with a most probable velocity offset of about -350 km s$^{-1}$ relative to
the systemic redshift. For the highest column densities, corresponding
to a covering fraction C$_f <$0.5, the line profile begins to show a
``flat core'', which is not favored by the data.  Therefore, it appears
that the best fitting models favor a relatively high covering fraction. We
note that the best fit Doppler parameter and velocity offset are not
very sensitive to the range of columns and covering fractions explored. We will further explore the wind properties in \S\ref{ssec:wind}.

{
\section{The origin of the line emission}
\label{sec:diagnostics}

The line emission of interstellar gas can be powered by a number of
different astrophysical processes like energetic photons from
starbursts and AGN, cosmic rays, or mechanical energy injected into
the ISM by stars, AGN, or various dynamical processes. We will now
analyze the emission lines of the molecular and ionized gas in 3C~326~N
to identify the physical mechanism that is giving rise to the luminous
line emission in this galaxy. 
%Before investigating for each of these
%processes whether it has a major contribution to heating the gas in
%3C~326~N, we will first focus on the H$_2$ line emission using a new
%%diagnostic diagram based on the CO(1-0), H$_2$ S(0) to S(3) and PAH
%7.7$\mu$m emission (Figure~\ref{fig:moldiag}). At the end of this
%section we will also analyze the optical emission line spectrum to
%infer the heating mechanism of the optical emission-line gas.

\subsection{Molecular emission-line diagnostics}
\label{ssec:moldiag}

Emission-line ratios in the optical are a popular tool to
investigate the heating mechanism of the ionized gas
\citep[e.g.,][]{baldwin81,VO87}, and to differentiate between
star-forming galaxies and AGN. Somewhat in analogy we will now
construct a molecular diagnostic diagram based on the CO(1-0),
pure-rotational H$_2$ and PAH emission shown in
Figure~\ref{fig:moldiag}. We start with a short discussion of the
astrophysical origin of each tracer before presenting the resulting
diagram.

Critical densities for the S(0) to S(3) lines are moderate \citep[10
to $10^4$ cm$^{-3}$ in molecular gas at 500 K;][]{lebourlot99}, so
that the lowest rotational states of H$_2$ are populated by
collisions rather than fluorescence. The first rotational lines of
H$_2$ are therefore dominant cooling lines for molecular gas over a
wide temperature range of $\sim 10^{2-3}$ K. They measure the power
dissipated by all heating processes of the warm molecular gas (T $>$
100 K).  This includes UV or X-ray photons produced by young stars in
photon or X-ray dominated regions, or the dissipation of mechanical
energy as proposed by \citet{guillard09} for Stephan's
Quintet. \citet{ferland08} proposed that cosmic rays could be the dominant 
energy source powering the extended warm intracluster H$_2$
line in the Perseus galaxy cluster. The CO(1-0) emission-line luminosity is associated with the
cold molecular phase, to the extent where it is commonly used to
estimate the cold molecular gas mass \citep[e.g.,][]{solomon97}. {\it
Thus, if the gas traced by CO and H$_2$ is physically associated, then
the H$_2$ to CO(1-0) ratio is a measure of the total heating of the
molecular gas per unit molecular gas mass.}

PAH emission in star-forming regions is found along the surfaces of
molecular clouds heated by UV photons, but not within HII regions,
where PAHs are destroyed \citep[e.g.][]{cesarsky96,tacconi05}. In
addition, on galactic scales, some of the PAH emission may originate
from the diffuse interstellar medium rather than star-forming
regions}, as argued by \citet{draine07b} based on the modeling of the
dust emission of SINGS galaxies. In either case the PAH emission is
powered by UV photons. {\it Thus, we can use the ratio of PAH to H$_2$
  emission as an empirical measure for the contribution of UV photons
  to the total H$_2$ heating.} This is supported by the tight
correlation between the 7.7$\mu$m PAH feature and H$_2$ luminosity
found in star-forming regions \citep{rigopoulou02, roussel07}.

In Figure~\ref{fig:moldiag} we combine both line ratios into one
diagnostic diagram.  We also include the position of 3C~326~N,
Stephan's Quintet (SQ), and of several H$_2$ -luminous radio galaxies
from the sample of \citet{ogle09} which have published CO
measurements. We exclude central galaxies in cooling-flow clusters
where CO line emission is extended over sizes much greater than the
IRS slit.  We also show star-forming galaxies and AGN from the SINGS
survey, taken from \citet{roussel07}. To convert 8$\mu$m flux
densities into PAH 7.7$\mu$m luminosities, we used the prescription of
\citet{roussel07}. We used the starburst template of \citet{brandl06}
to bring the integrated PAH fluxes of the radio galaxies measured by
\citet{ogle09} on the same scale. In Figure~\ref{fig:moldiag},
3C~326~N and the SQ shock are the most extreme representants of H$_2$
luminous galaxies with the highest ratio of H$_2$-to-PAH emission and
H$_2$-to-CO emission.

\subsection{What powers the H$_2$ emission?}

\subsubsection{Energetic photons}
\label{ssec:photons}
We will now use our molecular emission-line diagnostics to show that
UV photons produced in PDRs cannot make a major contribution to
heating the molecular gas in 3C~326~N.

In the diagnostic diagram shown in Figure~\ref{fig:moldiag}, black
lines mark the line ratios derived for PDR models to illustrate in
which parts of the diagram the H$_2$ heating is dominated by UV
photons. The H$_2$ line emission in PDRs is derived from the
calculations of \citet{kaufman06}. The total extinction of the PDR
models is A$_V=$10 mag, large enough to ensure that most of the
incident UV light is being absorbed. For the PAH
emission, we assumed that $\nu I_{\nu,8}=0.16\times F_{IR}$, where
$F_{IR}$ is the bolometric infrared flux, and $I_{\nu,8}$ the
flux density at 8$\mu$m. This conversion factor corresponds to the
models of \citet{draine07a} for a PAH-to-dust mass fraction $q_{PAH} =
3.55\%$ \citet{draine07a}, the median value for galaxies with Galactic
metallicity in the SINGS sample \citep{draine07b}.

The star-forming galaxies from SINGS are our control sample with
'ordinary' line emission. They fall into the portion of the diagram
spanned by the PDR models, as expected if most of their molecular line
emission is powered by star formation. Their positions in the diagram
suggest a ratio between the intensity of the UV field and the gas
density of about 0.1 to 1 cm$^{-3}$, and a UV intensity several times
higher than that in the solar neighborhood.

In galaxies which do not fall into the portion of the diagram spanned
by the PDR models, the molecular gas is heated by another mechanism
than photons produced in star-forming regions. This includes 3C~326~N,
3C~424, and the intergalactic shock in Stephan's Quintet as the most
extreme cases, as well as several other H$_2$ luminous radio
galaxies. Similarly, about half of the SINGS AGN fall into this
region, albeit not at its extreme end.

For the sake of completeness, we should note that X-ray heating
seems also unlikely to explain the H$_2$ line emission in the SINGS
AGN \citep{roussel07}. Similarly, \citet{ogle09} find that the observed
X-ray luminosities of their sample of H$_2$ luminous radio galaxies
are not sufficient to power the H$_2$ line emission.
\subsubsection{Cosmic rays}
\label{ssec:cosmicrays}

\citet{ferland08} recently proposed that heating by cosmic rays may
explain the extended, filamentary emission of warm molecular hydrogen
in massive galaxy clusters experiencing episodes of AGN feedback from
their massive central galaxy. We discuss this excitation mechanism
based on an approach that is broadly similar to that of \citet{ogle09}

The critical density of the S(0) and S(1) lines is sufficiently low
\citep{lebourlot99} to allow us to safely assume that the J=2 and 3
levels of H$_2$ are thermalized by collisions. The S(0) and S(1) line
fluxes in \citet{ogle09} correspond to an emission-line luminosity of
$1.9 \times 10^{41} \, {\rm erg \, s^{-1}}$. Dividing this luminosity
by the H$_2$ mass of $9.5 \times 10^8 \, M_\odot $ from
\citet{ogle09}, we can estimate a cooling rate through the H$_2$ S(0)
and S(1) lines of $3.4 \times 10^{-25} \, {\rm erg \, s^{-1} \,
  H_2^{-1}}$. Including the S(2) and S(3) lines increases this
luminosity to $9 \times 10^{-25} \, {\rm erg \, s^{-1} \,
  H_2^{-1}}$. What ionization rate from cosmic rays would be required
to provide the corresponding amount of heat?

Gas heating associated with the ionization of H$_2$ by cosmic rays is
$\sim 12~$eV per ionization, including the contribution from
H$_3^{+}$ recombination and H$_2$ re-formation
\citep[e.g.][]{lepetit06}. To balance line cooling, the ionization
rate per H$_2$, $\zeta_{H2}$, must be few $\times 10^{-14} {\rm
  s^{-1}}$.  For such a rate, cosmic rays are the main destruction
path of H$_2$ molecules. The molecular gas fraction depends on the
rate of ionization and gas density $\zeta_{H2} /n_{H2}$, and the gas is
molecular for n$_H > 10^4$ cm$^{-3}$ \citep[see Figure 2 of][]{ferland08}. At such
densities, the H$_2$ rotational states are thermalized up to $J \ge
5$, and the higher-J lines up to S(3) are also cooling lines.
Dividing the H$_2$ S(0)-S(3) luminosity by the energy released per ionization, we
estimate the required ionization rate, $\zeta_{H2} \sim 5 \times
10^{-14} {\rm s^{-1}}$. This value is larger than what is inferred
from H$_3^+$ observations in the Milky Way, by a factor 100 for the
diffuse interstellar medium \citep{indriolo07}, and by more than a
factor 10 for the molecular gas within 200 pc from the Galactic center
\citep{goto08}. It is therefore unlikely that the warm molecular gas
in 3C~326~N is powered by cosmic rays.

\subsubsection{Mechanical heating through shocks}
Having ruled out UV and X-ray photons as a possible mechanism to power
the line emission in 3C~326~N, and having shown that cosmic rays are
unlikely, our best remaining candidate is mechanical heating through
shocks. This is in close analogy to the model of \citet{guillard09}
who studied the intergalactic shock in Stephan's Quintet, and found
that the H$_2$ line emission is powered by the dissipation of kinetic
energy injected by the interaction of two galaxies. 3C~326~N and Stephan's
Quintet fall into similar regions of our diagnostic diagram
(Figure~\ref{fig:moldiag}), suggesting their line emission is powered
by a similar physical mechanism.
In \S\S \ref{ssec:massbudget} and
\ref{sec:enmom} we will demonstrate that the radio source is the only
viable energy source which can explain all of the phenomenology
observed in 3C~326~N, and present a physical framework through which
the energy injection by the radio source and the molecular line
emission can be related.

\subsection{What powers the optical line emission?}
\label{ssec:optdiagnostics}

In the previous section we presented new, 'molecular'
diagnostics to argue that the luminous H$_2$ emission in 3C~326~N (as
well as other H$_2$ radio galaxies) may be powered by the dissipation
of mechanical energy in the interstellar medium of the host galaxy. It
may be illustrative to use our SDSS spectra to compare this result
with the optical diagnostics.

Table~\ref{tab:3c326emlines} lists the fluxes measured for various
optical emission lines in 3C~326~N. Comparison with the classical
BPT-diagrams \citep{baldwin81,VO87,kewley06} shows that 3C~326~N falls
within the LINER \citep{heckman80} part of the diagrams, similar to
most H$_2$ luminous radio galaxies of \citet{ogle09} and also the
galaxy-wide shock in Stephan's Quintet \citep{xu03}.

The excitation mechanism of the atomic lines in LINERs is a
long-standing issue in the literature where some studies favor
photo-ionization by the AGN \citep[e.g.,][]{VO87} while other studies
propose shock excitation
\citep[e.g.,][]{clark98,dopita97,monreal06}. \citet{dopita95,dopita96}
show that the spectral characteristics of LINERS can be modeled with
excitation from fast, radiative shocks without the emission
from the radiative precursor of the shock. 
The radiative precursor of a fast shock moving into low-density gas
adds an emission component which has the spectral characteristics of
photoionized gas with a high ionization parameter. However, the line
ratios in 3C~326~N, in particular the ratio between the [NeII]/[NeIII]
lines in the mid-infrared, suggest a low ionization parameter. This
apparent contradiction can  be resolved if a clumpy or filamentary gas
distribution enhances the 'leaking' of hard photons along sight lines where the photons can escape without interacting with the pre-shock gas. As in the shock
of Stephan's Quintet, most of the gas in 3C~326~N is in H$_2$ at
densities $n_H> 10^3$ cm$^{-3}$ much higher than the density of the
ionized gas \citep{guillard09}, in good agreement with this scenario.

Quantitatively, using the models of \citet{allen08} for 3C~326~N, we
find that the ratios between the main optical lines in Table
\ref{tab:3c326emlines} are close to those of a 250 km s$^{-1}$ shock
moving into ionized gas with a 1 cm$^{-3}$ preshock density and a low
magnetic parameter, B$/\sqrt{n} <$1 $\mu$G cm$^{-3/2}$.  Observed and
modelled line ratios are compared in Table \ref{tab:optshock}. This is
not a unique solution, in particular, the line ratios are also
compatible with lower densities for similar shock velocities.  A
detailed analysis of the alternative models \citep[and model
  parameters, as done in ][]{dopita97} for the interaction between
radio jet and interstellar medium in M87 is beyond the scope of this
paper.  

We will merely use these models to determine the bolometric luminosity
of the ionized gas from the observed H$\beta$ luminosity. Using the
output of \citet{allen08} for shock velocities below about 300 km
s$^{-1}$ we can calculate the ratio of H$\beta$ luminosity, ${\cal
  L}_{H\beta}$, to the total luminosity produced by the shock. We find
that ${\cal L}_s \sim 200\pm 50 {\cal L}_{H\beta}$. This bolometric
correction is slightly larger for magnetic parameters below the
equipartition value.  This large correction factor is due to the much
larger brightness of the UV emission lines relative to the optical
line emission, which we did not observe. With the measured H$\beta$
flux and using this model at face value, we find ${\cal L}_{s} \sim
1\times 10^{43}$ ergs s$^{-1}$, a factor $\sim$10 more than the total
H$_2$ line luminosity in the S(0)-S(7) rotational lines
\citep{ogle09}. This corresponds to the isobaric cooling of gas at
temperatures of $\sim 10^6$ K for a mass flow of order of a few 100
M$_{\odot}$ yr$^{-1}$. This mass flow is very large, and we will
discuss in \S\ref{ssec:dissipationtime} how this result may be
interpreted as a repeated mass cycling between gas phases driven by
interactions between the radio jet and the multiphase interstellar
medium of the host. Obviously, due to the uncertainties in the
measurement and modeling all of these values have uncertainties of
factors of a few.

It is well possible that not all of the optical line emission is
excited mechanically. The dynamical interaction between gas phases
must include turbulent mixing between the cold and hot gas, which
produces extreme UV radiation along the surfaces of cold clouds. This
radiation is an additional, and maybe significant, source of
ionization powered by the thermal energy of the
cocoon. \citet{crawford92} show that this mechanism could account for
parts of the line emission in ionized gas in cooling flows, consistent
with observed line ratios.  In this case, the bolometric correction
with respect to H$\beta$ depends on the effective temperature of the
gas after mixing and could be significantly smaller.  With
regard of this uncertainty on the radiative excitation, we will in the
following adopt a fiducial value of $1\times 10^{43}$ erg
s$^{-1}$. Again, the uncertainties of this estimate are likely factors
of a few.

\subsection{Photoionization by stars and the AGN}
\label{ssec:starsAGN}
In the above estimates we assumed that all of the optical line emission is due
to shocks, and that other processes like photoionization from star formation
or the AGN can be neglected. With the low star-formation rate and faint AGN
X-ray emission from 3C~326~N this galaxy is ideally suited to set stringent
limits on the contribution of these processes to the observed line
emission. For example, the observed star-formation rate of $\le 0.07$
M$_{\odot}$ yr$^{-1}$ would imply an H$\alpha$ luminosity ${\cal
 L}$(H$\alpha$)$\sim 1\times 10^{40}$ erg s$^{-1}$ (corresponding to a
H$\beta$ luminosity of few $\times 10^{39}$ erg s$^{-1}$) for a continuous
star-formation history and using the models of \citet{bc03}. This corresponds
to $\sim 5$\% of the measured H$\alpha$ luminosity of 3C~326~N.

We can also rule out a dominant role of the AGN in photoionizing the
gas, for the simple reason that the [OIII]$\lambda$5007 luminosity
alone exceeds the luminosity emitted in the X-ray. \citet{ogle09} find
an X-ray luminosity of $4\times 10^{40}$ erg s$^{-1}$ for 3C~326~N,
whereas our [OIII]$\lambda$5007 measurements indicate an
[OIII]$\lambda$5007 luminosity of $\log{\cal L}_{[OIII]} = 7\times
10^{40}$ erg s$^{-1}$ (including the extinction correction of a factor
2).  \citet{heckman04} find that the bolometric luminosity of quasars
scales with [OIII]$\lambda$5007 emission-line luminosity as ${\cal
  L}_{bol}\sim 3500 {\cal L}_{[OIII]\lambda5007}$. For 3C~326~N the
observed [OIII]$\lambda$5007 flux implies a bolometric luminosity of
few $\times 10^{44}$ erg s$^{-1}$, three orders of magnitudes greater
than that estimated from the X-ray measurement. This indicates that
the AGN photoionization is unlikely to be the dominant mechanism in
exciting the optical line emission in 3C326~N.

\section{Mass and energy budgets}
\label{ssec:massbudget}
 
The relative mass budgets of warm and cold molecular gas and the warm
ionized gas provide important constraints on the physical and
astrophysical conditions of a galaxy. By 'physical conditions' we
refer to the gas properties, which relate directly to the
observations. By 'astrophysical conditions' we refer to the galaxy
properties and the mechanism which is causing the observed gas
properties. Typically, in gas-rich, actively star-forming galaxies the
amount of cold molecular gas exceeds the masses of warm molecular and
ionized gas by factors 10$-$100 \citep[see][for samples of nearby
  galaxies and ULIRGs, respectively]{roussel07, higdon06}. We will now
show that this ratio is much smaller for 3C326~N.  All masses,
luminosities, and kinetic energies are summarized in
Table~\ref{tab:summary}.

\subsection{Molecular gas mass} 
\label{ssec:mol}

\subsubsection{Direct estimate of the luminosity and mass of warm molecular gas}
\label{sssec:warmmol}

The mass of warm molecular gas in 3C~326 has been estimated by
\citet{ogle07,ogle09} by fitting the H$_2$ S(0) to S(7) rotational
line fluxes with 2 or 3 components at different temperatures where
H$_2$ excitation and ortho-to-para ratios are assumed to be
thermalized. This yields a mass of $1.0\times 10^9$ M$_{\odot}$.

In the present analysis, we associate the H$_2$ emission with the dissipation of 
kinetic energy in the molecular gas. Therefore we follow a different approach
to estimate the mass and luminosity of warm molecular gas. 
As in \citet{guillard09}, we model the dissipation process with magnetic
shocks in molecular gas, which maximizes the H$_2$ luminosity per total
emitted power. In this sense, our results represent a lower limit to the
dissipated energy required to account for the observed H$_2$ luminosity. 

Each shock model includes a range of gas temperatures which depend on the
shock velocity, pre-shock gas density and intensity of the magnetic field.  We
use the grid of models presented by \citet{guillard09} for proton 
densities of n$_H = 10^3$ cm$^{-3}$ and $10^4$ cm$^{-3}$, respectively, an initial
ortho-to-para ratio of 3 and a magnetic parameter, $\rm B/\sqrt{n_H} = 1 \mu$G
cm$^{-3/2}$.

The shock velocity is the only parameter that we allow to vary.  A combination
of three shocks is required to match the emission in all 8 H$_2$ lines, S(0)
to S(7). These fits provide a scaling factor for each of the three shocks
which represents a mass flow, the amount of gas traversing the shocks per unit
time.  As discussed in \citet{guillard09}, this fit is not unique but the
proposed solution may be used to quantify the relevant range of shock
velocities, and to estimate the warm gas masses by multiplying the mass flows
with the gas cooling time, and the shock luminosities by integrating over all
cooling lines. 

\begin{figure}
\centering
\includegraphics[width=0.5\textwidth,angle=90]{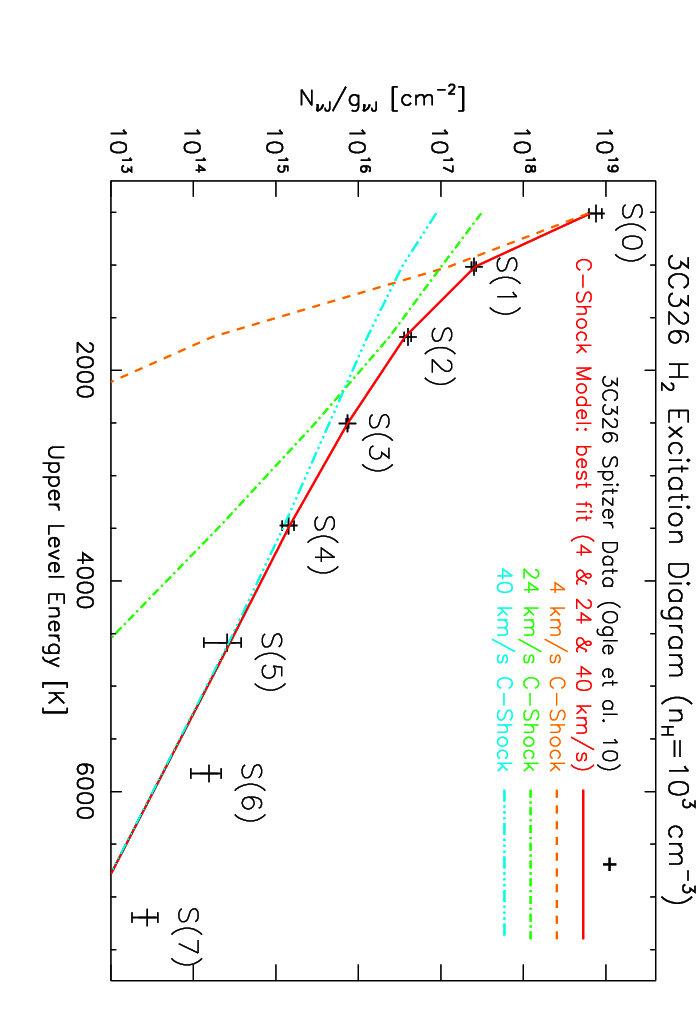}\\
\includegraphics[width=0.5\textwidth,angle=90]{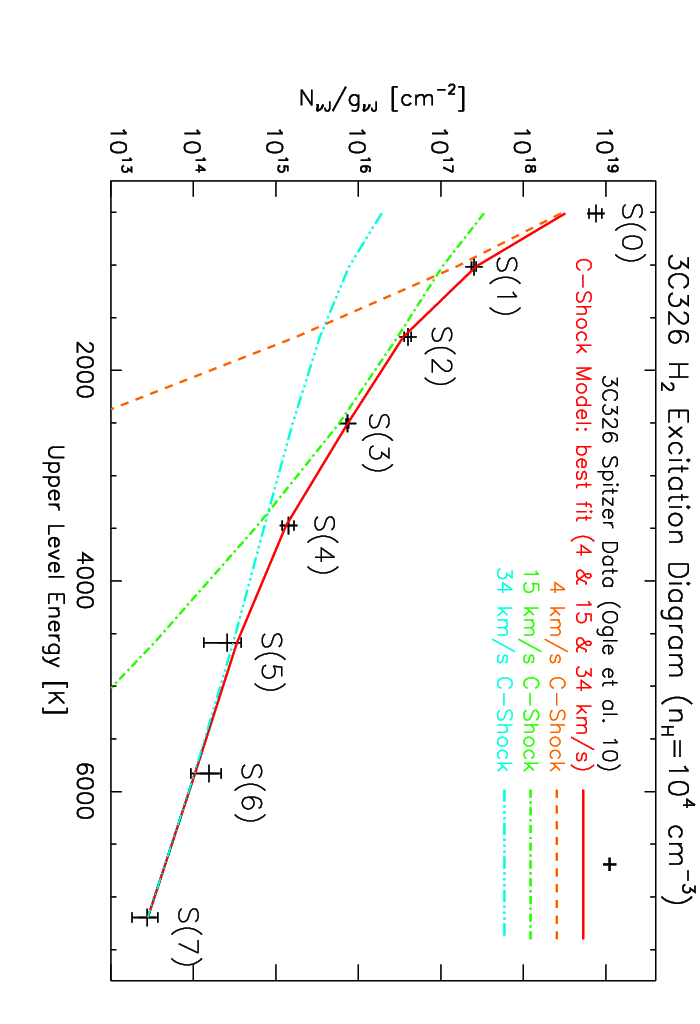}
\caption{Observed H$_2$ excitation diagrams for 3C326 (1$\sigma$ error
  bars) and model results of magnetic shocks \citep[see][for
    details]{guillard09}, for pre-shock densities of n$_H=10^3$ (upper
  panel) and $10^4$ cm$^{-3}$ (lower panel). Dashed lines show the
  contribution of each C-shock velocity to the best-fit model. The red
  line shows the combination of these 3 shocks.}
 \label{fig:excitationdiagrams}
\end{figure}

\begin{figure}
\centering
\includegraphics[width=0.45\textwidth]{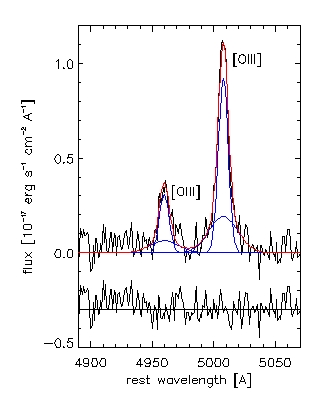}
\includegraphics[width=0.45\textwidth]{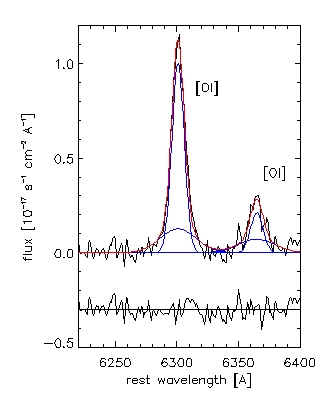}
\caption{Multiple-component fits to the
  [OIII]$\lambda\lambda$4959,5007 and [OI]$\lambda\lambda$6300,6363
  line emission in 3C~326~N. The broad component is best seen in the
  [OIII]$\lambda$5007 emission line. We assumed the same redshifts and
  line widths for all lines.}
 \label{fig:optlinefits}
\end{figure}

\begin{figure}
\centering
\includegraphics[height=0.45\textheight]{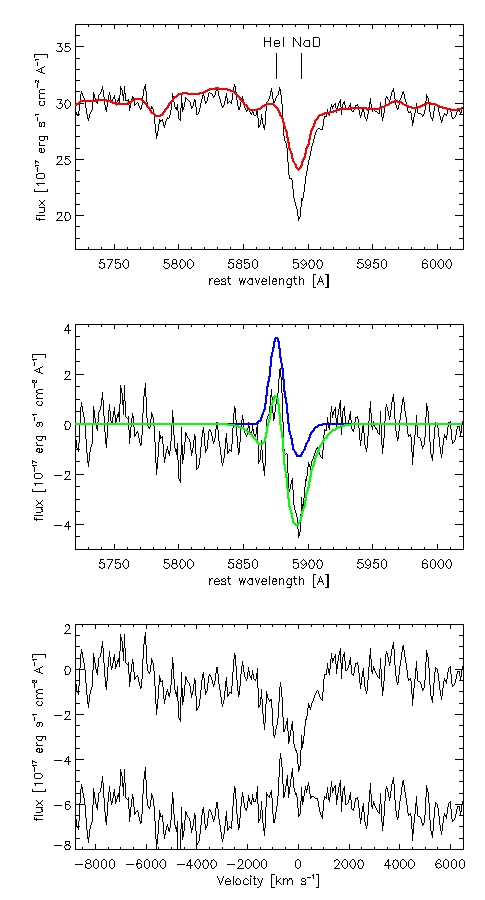}
\includegraphics[height=0.45\textheight]{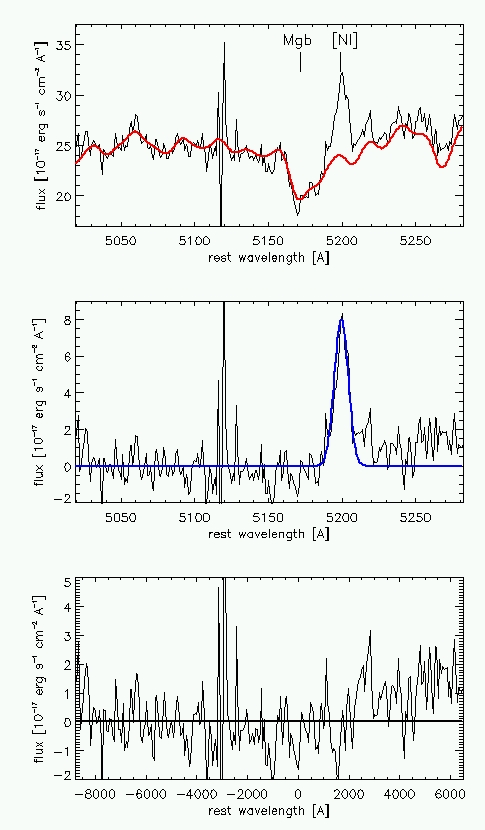}\\
\caption{{\it left:} NaD absorption line in 3C~326~N. In the upper
  panel, the spectrum is shown in black, and the red line shows the
  stellar continuum for our best stellar population fit. The two NaD
  components are not resolved at R$=$1800. In the mid panel, we show
  the spectrum with the best-fit stellar continuum subtracted. 
    The green line in the mid panel marks our absorption line fit with
    HeI$\lambda$5876, and two Gaussian NaD components, one at the
    systemic velocity, another one with an offset of -350 km s$^{-1}$
    representing the blue wing. The blue line shows the HeI line and
    the stellar NaD component. In the lower panel, we show the
  spectrum with the stellar continuum and the HeI emission-line
  subtracted (upper spectrum) and the residual after subtraction
    of HeI, the stellar component, and the NaD absorption line
    corresponding to our best-fit wind model. The residual spectrum
  is shifted by an arbitrary amount along the ordinate. The abscissa
  gives the velocity offsets relative to a systemic redshift of
  z=0.090. The Na D profile shows evidence for absorption with
  blueshifts of up to $\sim -1800$ km s$^{-1}$, and potentially, at
  lower signal-to-noise levels, of up to $-6000$ km s$^{-1}$. {\it
    right} The same for Mgb, a line which does not have a strong
  interstellar component. The residual (lower panel) is consistent
  with noise after subtraction of the stellar component and the
  NI$\lambda$5199 line.}  
\label{fig:3C326NaDspec}
\end{figure}

\begin{figure}
\centering
\includegraphics[width=0.7\textwidth]{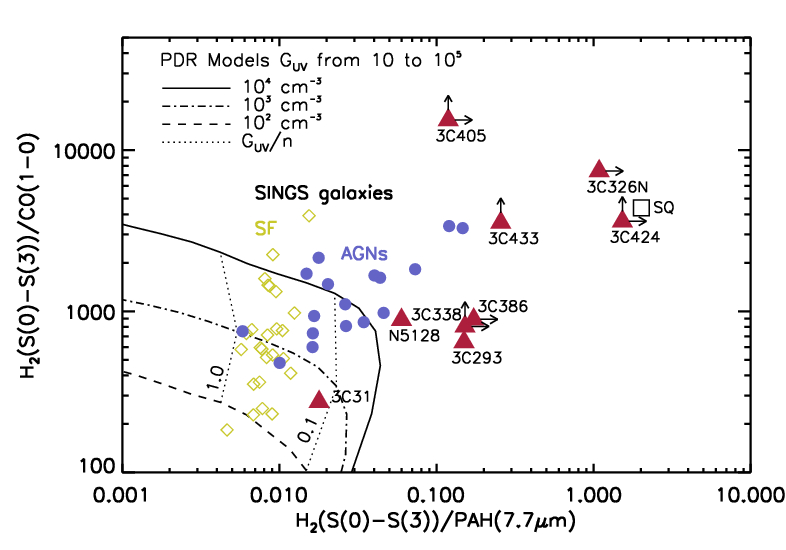}\\
\caption{Molecular diagnostic diagram to distinguish between AGN and
  star formation based on PAH bands, CO(1-0) and mid-infrared H$_2$
  line emission. Lines mark PDR models with different assumptions for
  the UV radiation fields. Star-forming galaxies from the SINGS survey
  fall within the portion of the diagram spanned by PDR models (black
  lines), unlike Stephan's Quintet, 3C~326~N and a number of other H-2
  luminous radio galaxies with published CO(1-0)
  observations. Consequently, the H$_2$ line emission in these
  galaxies cannot be powered by energetic photons produced in
  star-forming regions. Values for the SINGS galaxies are taken
    from \citet{roussel07}. H$_2$ and PAH fluxes for the H$_2$
    luminous radio galaxies are taken from \citet{ogle09}, who show
    the relationship between H$_2$-to-PAH 7.7$\mu$m versus 24$\mu$m
    continuum luminosity.} \label{fig:moldiag}
\end{figure}

Excitation diagrams for models with two different gas densities,
$10^3$ cm$^{-3}$ and $10^4$ cm$^{-3}$, respectively, are shown in
Fig.~\ref{fig:excitationdiagrams}. The corresponding H$_2$ line fluxes
are listed in Table~\ref{tab:shock_H2_fluxes_3C326}. They are smaller
than the kinetic energy fluxes one may compute from the mass flow and
shock velocity because some of the energy is transferred to the
magnetic field. The luminosities of the mid-infrared H$_2$ lines are
close to the bolometric values obtained by integrating the emission in
all lines because the H$_2$ rotational lines are the main cooling
lines of magnetic shocks. Summing over the three shocks we estimate a
total luminosity of $10^{42}$erg s$^{-1}$ for the molecular Hydrogen.

Table~\ref{tab:shock_mass_3C326} lists the gas cooling times and
H$_2$ masses down to temperatures of T$=$150 K for each of these models.
The total mass we obtain for gas at temperatures larger than T$=$150 K,
$1.3-2.7 \times 10^9 M_\odot$, is slightly larger than that derived by
\citet{ogle09}, and we will in the following assume a fiducial molecular gas
mass of $2\times 10^{9}$ M$_{\odot}$. The small difference between the
estimates of \citet{ogle09} and our results is due to different
ortho-to-para ratios and moreover, in our lowest-density models, the
S(0) and S(1) lines are not fully thermalized.

\subsubsection{CO(1--0) luminosity and total molecular gas mass}
\label{sssec:coldmol}

The CO emission-line luminosity is often used as an empirical measure of the
mass of cold molecular gas.
In \S\ref{ssec:coin326}
we estimated an integrated CO(1-0) emission-line flux of I$_{CO}$=1.0$\pm$0.2
Jy km s$^{-1}$ from our millimeter spectroscopy at the IRAM Plateau de Bure
Interferometer, extracted from a 5\arcsec\ aperture. Using Equation~3 of
\citet{solomon97} we translate this value into a CO(1--0) emission-line
luminosity of $L_{CO}\prime=$3.8$\times 10^8$ K km s$^{-1}$ pc$^2$ at a
redshift of z=0.090.  Applying a Galactic H$_2$-to-CO conversion factor of
$4.6 M_\odot/(\rm K\ \rm km\ \rm s^{-1} \rm pc^2) $ as determined for gas in
the molecular ring of the Milky Way \citep{solomon92}, this would correspond
to a mass of cold molecular gas of M$_{cold}^{326~N}= 1.5\times 10^9$
M$_{\odot}$. We will further discuss the appropriateness of this H$_2$
conversion and related uncertainties in \S\ref{sec:SF}.

Comparison with \S\ref{sssec:warmmol} shows that the amount of warm
molecular gas measured directly from the mid-infrared lines is similar
to the H$_2$ molecular mass inferred from the CO(1--0) flux estimated
from our PdBI observations.  This is a highly unusual finding compared
to star-forming galaxies where the ratio of warm to cold molecular gas
mass is of order $10^{-1}$ to $10^{-2}$ \citep{roussel07,higdon06} for
the same CO-to-H$_2$ conversion factor. However, it is not very
different from the ratio of $\sim 0.3$ found in molecular clouds near
the Galactic center \citep{rodrigues01}. 

The Galactic center may be an
adequate nearby analog for the properties of the molecular gas in
3C~326~N. In the Galactic center cold dust temperatures indicate that
the gas cannot be heated by UV photons \citep{lis01}, but possibly by
shocks or cosmic rays \citep{rodrigues01,yusefzadeh07}. The H$_3^+$
observations presented by \citet{goto08} imply an H$_2$ ionization
rate lower than the value required to account for the temperature of
the warm molecular gas of $\sim 150~$K \citep{rodrigues01} with
heating by cosmic rays, $\zeta_{H2} \sim 10^{-13} {\rm s^{-1}}$
\citep{yusefzadeh07}.

 From detailed
studies of several molecular species in the Galactic center,
\citet{lis01,huttemeister98} conclude that all of the molecular gas in
this environment may be warm at temperatures above T$=$50 K. By
analogy we caution that the usual distinction between cold and warm
gas measured through CO line emission and infrared H$_2$ lines,
respectively may not apply to 3C~326~N. Most of the CO line emission in
3C~326~N could in fact be associated with the warm gas. Observations of
higher-J transitions are necessary to test this hypothesis. We used
the RADEX LVG code \citep{vandertak07} to verify that the intensity of
the CO(1-0) line is relatively insentitive to the gas temperature for
a given CO column density. For example, in a temperature range of
  10-100 K, the intensity changes by about a factor 2 \citep[see
    also][]{lada96}.

Obviously, given these considerations, the standard conversion factor may not
apply at all. With these caveats in mind, we will in the following assume that
the two tracers do not represent the same gas, which would correspond to a
total gas mass of at most $3.5\times 10^9$ M$_{\odot}$. 

\subsubsection{Warm ionized gas mass}
\label{sssec:warmion}
We can also estimate the mass of ionized gas in 3C~326~N from our
SDSS spectra. Assuming case B recombination and following \citet{osterbrock89}
we can estimate an ionized gas mass, M$_{HII}$ from the H$\alpha$
emission-line luminosity, ${\cal L}_{H\alpha}$, by setting

\begin{equation}
M_{HII} = \frac{{\cal L}_{H\alpha}\ m_p } {h\ \nu_{H\alpha}\ \alpha^{eff}_{H\alpha}\
n_e} = 9.73 \times 10^6\ {\cal L}_{H\alpha,41}\ n^{-1}_{e,100}\  M_{\odot},
\label{eqn:hiimass}
\end{equation}
where $m_p$ is the proton mass, $h$ is Planck's constant,
$\alpha^{eff}_{H\alpha}$ and $\nu_{H\alpha}$ 
are the effective recombination cross section and the frequency of H$\alpha$,
respectively. $n_{e,100}$ is the electron density in units of 100
cm$^{-3}$. The H$\alpha$ luminosity, ${\cal  L}_{H\alpha,41}$, is given in
units of $10^{41}$ erg s$^{-1}$. 

In order to estimate extinction-corrected, intrinsic H$\alpha$
luminosities, we measure the H$\alpha$ and H$\beta$ fluxes from our
SDSS spectra and compare with the expected Balmer decrement of
H$\alpha$/H$\beta$=2.9. We find A$^{326~N}_{H\beta}$=0.8 mag.  This
implies an extinction-corrected H$\alpha$ luminosity of ${\cal
 L}(H\alpha)_{ext}^{326~N}= 2.8\times 10^{41}$ erg s$^{-1}$.

We also need to constrain the electron densities in the optical emission-line
gas, which we estimate from the line ratios of the density-sensitive
[SII]$\lambda\lambda$6716,6731 emission line doublet. Fitting each line of the
doublet with Gaussian profiles, we find a line ratio of R$^{326N}$ = 1.3. This
is near the low-density limit, and assuming an electron temperature of 10$^4$
K, these line ratios correspond to electron densities n$_e^{326}$=130
cm$^{-3}$. With these extinction-corrected H$\alpha$ luminosities and electron
densities, and using Equation~\ref{eqn:hiimass}, we estimate ionized gas
masses of M$_{HII}^{326}=2\times 10^7$ M$_{\odot}$.

\subsection{Kinetic energy provided by the radio source}
\label{ssec:radiojet}
3C~326 has a powerful, large radio source of Mpc size and resides in a
relatively low-density environment compared to powerful radio galaxies
generally \citep[][]{stocke79,willis78}.
Estimating the intrinsic properties of radio jets, in particular their
kinetic power and lifetimes, remains a challenge. The synchrotron
emissivity of the radio jet depends on the intrinsic jet power and the
ambient conditions, and the jet kinetic power is therefore not easily
derived. Estimates based on theoretical arguments suggest factors of
$\sim$10$-$1000 between radio luminosity and kinetic power
\citep[e.g.][]{young93,bicknell97}.

Given these uncertainties, it may be best to use empirical estimates
of the jet kinetic power. \citet{birzan04,birzan08} estimated the
kinetic power of radio sources by comparing the necessary energy to
inflate X-ray cavities in galaxy clusters with the luminosities of the
radio sources that are inflating them. They find a factor $\sim$100
between kinetic and bolometric radio luminosity, with a large upward
scatter of up to factors of a few 1000, perhaps suggesting that the
correlation is in fact a lower envelope. Using their scaling with the
radio luminosity measured at 327 MHz, and measured radio flux, we find
kinetic luminosities of ${\cal P}_{k,jet,327}^{326} = 4\times 10^{44}$
erg s$^{-1}$. For fluxes measured at 1.4 GHz, we find ${\cal
P}_{k,jet,1400}^{326} = 1.3\times 10^{45}$ erg s$^{-1}$.  Using the
estimate of \citet{merloni07} instead, which relies on the 5-GHz core
radio power, we find $2\times 10^{44}$ erg s$^{-1}$. \citet{willis78}
suggest a total energy content of $6\times 10^{59}$ erg in the radio
lobes, which would correspond to a kinetic luminosity of
few$\times 10^{44-45}$ erg s$^{-1}$ for a fiducial lifetime of the
radio source of $10^{7-8}$ yrs, in good agreement with the previous
estimates. The estimates derived with each method change,
reflecting the uncertainties of each approach, but the overall result
of this section holds, namely, that the kinetic luminosity is of order
few$\times 10^{44-45}$ erg s$^{-1}$.

A closely related quantity is the lifetime of the radio source. Jet
lifetimes may either be estimated based on radio spectral indices or
estimates of the velocity with which the jet expands, the former
typically giving significantly smaller values. For 3C~326
specifically, \citet{willis78} carefully investigated the
multi-wavelength radio properties finding spectral ages of $1-2\times
10^7$ yrs. Using their estimate of the Alfven speed and size estimate
with our cosmology would yield a kinetic age of $6\times 10^7$ yrs for
most parts of the radio source, although a region that is somewhat
spatially offset and has a different polarization angle, may be as old
as $\sim 2\times 10^8$ yrs \citep{willis78}. It is unclear whether
this is due to an extended activity period, several activity
outbreaks, or the second galaxy of the system, 3C~326~S, which also
has a radio core. At any rate, the detection of millimeter continuum
emission from 3C~326~N and 3C~326~S (\S\ref{ssec:co}, see also 
Fig.~\ref{fig:CO_map}) implies on-going activity in both nuclei.

\subsection{Kinetic energy of the gas}
\label{ssec:gaskinematics}
We will now give rough estimates of the related kinetic energies in
each gas phase $i$. The total kinetic energy estimated from the line widths is given by
E$_{\rm r,tot}= 3/2\ \Sigma\ m_i\ \sigma_i^2$. $m_i$ is the gas mass
in each phase, and $\sigma_i$ is the velocity dispersion for the
ensemble of clouds derived from the widths of the emission lines (and
given in Tables~\ref{tab:co1} and \ref{tab:3c326emlines} for CO and
the optical lines, respectively), where we set $\sigma$ = FWHM/2.355.
 
From the optical and millimeter spectroscopy we have direct
measurements of the line width, which is FWHM$\sim 600$ km s$^{-1}$
for the optical lines, and FWHM$\sim 350$ km s$^{-1}$ for the CO
lines. The rotational H$_2$ lines are not spectrally resolved with
IRS, yielding an upper limit of $\le$ FWHM$\sim 2500$ km s$^{-1}$,
which is not very constraining. Near-infrared observations of the
ro-vibrational lines would be an important test to infer the dynamical
state of the warm molecular gas. The line widths of CO and the optical
lines, which trace gas that is colder, and gas that is warmer than the
warm H$_2$, respectively, are not very different
(Figure~\ref{fig:CO_spectrum}). We therefore
expect that the line widths of warm H$_2$ will be in the same
range. 

Since these spectra are integrated, the line widths will be affected
by large-scale motion within the potential of the
galaxy. \citet{holt08} obtained longslit spectra of optical emission
lines in 16 nearby powerful radio galaxies, finding that the
kinematics are often very irregular and do not appear dominated by
rotation. Where they found regular velocity gradients resembling
rotation curves, their data suggest that within radii of 2.5 kpc
(corresponding to our aperture) we may sample a range of rotational
velocities less than 100$-$150 km s$^{-1}$.

For a measured line width of
FWHM$\sim$500$-$600 km s$^{-1}$ a velocity gradient of $\sim 150$ km
s$^{-1}$ subtracted in quadrature will lead to a negligible correction
of $\sim$25 km s$^{-1}$ (35 km s$^{-1}$ for the measured CO line
width of FWHM=350 km s$^{-1}$). Therefore we do not believe that 
gravitational motion will have a large impact on our
measurements. With the velocity and mass estimates given above and in
\S\ref{ssec:massbudget}, respectively, we obtain a total kinetic
energy of E$_{H2}^{326}= 3\times 10^{57}$ erg for the cold and warm
molecular gas.  The ionized gas has a much smaller mass and a
negligible kinetic energy of order $5\times 10^{55}$ erg. Kinetic
energies are summarized in Table~\ref{tab:summary}.

\section{A wind without a starburst}
\label{sec:outflow}

\subsection{Mass and energy loss rate of the neutral wind}
\label{ssec:wind}
In attempting to estimate the characteristics of the multiphase medium,
absorption lines can play an important role. In \S\ref{ssec:NaD} we described
the detection of a significant interstellar component to the Na~D absorption
line in 3C~326~N with a systematic velocity offset to the blue and a pronounced
blue wing. This is very similar to what is frequently found in starburst
galaxies \citep[][]{heckman00,martin05,martin06}, where blueshifted Na~D
absorption is commonly interpreted as strong evidence that galaxies with
intense star formation are driving energetic outflows.

Similarly for radio-loud AGN, \citet{morganti05} give compelling
evidence for outflows of neutral material based on
studies of HI absorption line profiles at 1.4 GHz. They identify
pronounced blueshifted components with velocities of up to $\sim$ 2000
km s$^{-1}$ in a number of galaxies, and a clear excess of blueshifted
relative to redshifted material. They emphasize that they find
blueshifted material in {\it all} radio galaxies with sufficiently
deep HI spectroscopy, suggesting that outflows of neutral gas may be
common amongst powerful radio-loud AGN.  Three galaxies observed with
Spitzer-IRS have also been observed in HI by \citet{morganti05}, and
interestingly the two which have the more pronounced HI absorption
have also luminous H$_2$ line emission \citep{ogle09,haas05b}, as
expected if outflows and H$_2$ line emission are physically related.
The relative velocity of the blue wing of the Na D line suggests a
terminal velocity of $\sim -1800$ km s$^{-1}$ for 3C~326~N, within the
range found by \citet{morganti05}.

To estimate the energy and mass loss rates of the outflow, we need to
constrain the associated column density of neutral gas. The
relationship between the column density of Hydrogen, N$_H$, and that
of Sodium, N$_{Na}$, is determined by the abundance of Na relative to
H and the ionization correction for Na (NaI and NaII). If, as is
common for starburst-driven winds \citep[e.g.,][]{heckman00}, we
assume a solar abundance ratio, a depletion factor of 10
\citep{morton75} or perhaps more \citep{phillips84}, an ionization
correction of a factor of 10, and a NaI column density of 10$^{14}$
cm$^{-2}$, we find a total H column density of N$_H$$\sim$10$^{21.7}$
cm$^{-2}$. To be conservative, we will adopt N$_H$$\sim$10$^{21}$
cm$^{-2}$ as a fiducial value. This opens the possibility that
3C~326~N may have a significant component of warm neutral gas, but
given the absence of HI spectroscopy, the exact amount of this gas is
difficult to quantify, in particular, since parts of the Na~I could be
associated with outflowing molecular gas.

To directly associate the kinematics of the gas with an outflow rate
of mass and energy, we assume a simple model of a mass conserving flow
with a constant velocity, which extends from some minimal radius to
infinity.  This gives mass outflow rates of:

\begin{eqnarray}
\dot{\rm M}_{\rm N_H} \sim 50\  C_f\ \frac{\Omega}{4\pi}\ \frac{r}{1 \rm kpc}\ \frac{N_H}{10^{21}\
  \rm cm^{-2}}\ \frac{v}{350\ \rm km\ \rm s^{-1}}\ {\rm M}_{\sun}\ {\rm yr}^{-1}, 
\end{eqnarray}
and  energy loss rates of
\begin{eqnarray} 
\dot{\rm E}_{\rm N_H} \sim 10^{42.5}\  C_f\ \frac{\Omega}{4\pi}\
\frac{r}{1 \rm{kpc}}\ \frac{N_H}{10^{21}\ \rm{cm}^{-2}}\ (\frac{v}{350\ \rm km\
  s^{-1}})^3\ {\rm erg}\ {\rm s}^{-1}, 
\end{eqnarray}
respectively, where $C_f$ is the covering fraction,
$\frac{\Omega}{4\pi}$ is the opening angle, $v$ is the outflow
velocity. Covering fraction, opening angle and column density are
degenerate. However, relevant for our analysis is the product of the
three quantities, so that this does not influence our results. But
note that our analysis can only be accurate at an order-of-magnitude
level.

We have argued that the NaD lines are likely associated with a
significant (but difficult to quantify without direct HI observations)
reservoir of warm atomic gas. The size of the (marginally) resolved CO emission
(2.5\arcsec$\times$2.1\arcsec) is likely similar to the size of the Na
D absorbing region.  If we take the geometrical mean of the half beam
width as the entrainment radius, $\sim$1.8 kpc, allowing for a
covering fraction of 0.5 to 1, and corresponding column densities of
10$^{21.3-21.7}$ cm$^{-2}$, an opening angle of the outflow of $\pi$,
and an outflow velocity of 350 km s$^{-1}$ corresponding to the offset
velocity found with our line fit (and ignoring the Doppler parameter
which suggests higher velocities), we find a mass outflow rate of
about 30-40 ${\rm M}_{\sun}\ {\rm yr}^{-1}$ and an energy loss rate of
$\sim$10$^{42-43}$\ ${\rm erg} {\rm s}^{-1}$.

In starburst galaxies, the energy loss rates estimated with the same
method and assumptions correspond only to a few percent of the
injected mechanical power \citep[e.g.][]{heckman00,martin05}. If the
same factor approximately applies for 3C~326~N, we would expect a
mechanical power of at least few $\times 10^{43}$ erg s$^{-1}$
(\S\ref{ssec:radiojet}).

\subsection{Outflow energetics}
\label{ssec:outflow}
The outflow properties of 3C~326~N are very reminiscent of what is observed
in local starburst galaxies \citep[e.g.][]{heckman00,martin05}, which
have column densities of $\sim$few$\times 10^{21}$ cm$^{-2}$, mass and
energy outflow rates of a few to a few 10s M$_{\odot}$ yr$^{-1}$ and
10$^{43}$ erg s$^{-1}$, respectively, and reach maximal blueshifts of
$\sim$400$-$600 km s$^{-1}$. AGN and starburst activity often coincide
making it difficult to uniquely identify the energy source that is
driving the wind. For 3C~326~N this is not the case. \citet{ogle07}
estimated an upper limit of SFR$\le$0.07 M$_{\odot}$ yr$^{-1}$ for
the total star formation rate.
Assuming that all of this star formation occurs within
regions covered by the 3\arcsec\ SDSS fibre, this would correspond
to a star-formation intensity of SFI$\le 0.004\ M_{\odot}$ yr$^{-1}$
kpc$^{-2}$. This is more than an order of magnitude lower than the
SFI$\sim 0.1$ M$_{\odot}$ yr$^{-1}$ kpc$^{-2}$ threshold found by
\citet{heckman03} for starbursts that drive winds. We come to a similar
conclusion when comparing with the supernova rate expected for constant
star formation with SFR$= 0.07$ M$_{\odot}$ yr$^{-1}$. Using \citet{bc03}
we find a supernova rate of $3.5\times 10^{-4}$ yr$^{-1}$. Assuming
that all of the ``canonical'' energy release of a supernova, $10^{51}$
erg, will thermalize, we find an energy injection rate of $10^{40}$
erg s$^{-1}$, about three orders of magnitude lower than the
mechanical power of the wind in 3C~326~N or the line emission of the
warm ionized and molecular gas. For more realistic thermalization
efficiencies of few tens of percent \citep{strickland09}, the energy
injection from supernovae will be even lower by factors of a few.

In addition, the terminal velocity of $-$1800 km s$^{-1}$ is
$\sim$3-4$\times$ greater than terminal velocities typically found
in starburst-driven winds \citep[which are of order few 100 km
s$^{-1}$][]{heckman00,martin05,martin06} and similar to some of the
most powerful starbursts (which have star-formation rates orders of
magnitude higher than 3C~326).  Both arguments indicate that the outflow
in 3C~326~N is {\it not} related to star formation. This velocity is
also significantly larger than what we may expect from velocities due
to a possible interaction with the nearby galaxy 3C~326~S \citep[see the
discussion in ][]{ogle07}. The only plausible candidate driving this
outflow is a radio-loud AGN.

\section{Energy and momentum exchange in the multiphase cocoon}
\label{sec:enmom}

We will now use our observational results to construct a scenario for the
interaction between jet and interstellar medium of the host galaxy, which
explicitly includes the molecular gas. We will argue that the kinematics and
emission-line luminosities of 3C~326~N are most likely related to the energy
injected by the radio jet, which is being dissipated by the multiphase
interstellar medium of the host galaxy. We describe and quantify the energy
flow and associated timescales. 

\subsection{Energy, mass and momentum flow}

Much theoretical effort has been dedicated to describing the interactions of
radio jets with the ambient medium \citep[for early studies see,
 e.g.,][]{scheuer74,begelman89}. Observational evidence that jets may have a
profound influence on the interstellar medium of their host galaxies has been
known for at least two decades \citep[e.g.,][]{vanbreugel85,pedlar85,
tadhunter91,eales93}.  The 'cocoon model' 
describes interactions between radio jet and ambient gas in form of a 'waste
energy basket' \citep{scheuer74} of hot, low-density, but high-pressure
material that surrounds the thin relativistic jet. This 'cocoon' will expand
into the ambient gas and may entrain matter ablated from denser clouds of
colder material \citep[e.g.,][]{begelman89}, drastically enhancing the
efficiency with which the jet interacts with the gas of the galaxy compared to
simple interactions along the jet working surface \citep['dentist's
 drill',][]{scheuer82}. We will now extend this scenario by explicitly taking
into account the energy, mass, and momentum exchange between different gas
phases which result from the injection of mechanical energy by the radio
source. This scenario represents the synthesis of our observational results
discussed in \S\S\ref{sec:diagnostics} to \ref{sec:outflow}.

Fig.~\ref{fig:flow_diagram} illustrates our basic scenario of the
energy and momentum flow in the cocoon. Part of the mechanical energy
injected by the radio jet is ultimately transformed into thermal
energy of the warm atomic and molecular gas, giving rise to the
observed line emission.  A part of the jet kinetic power is also
translated into thermal and bulk kinetic energy of the hot cocoon
plasma (first box in Fig.~\ref{fig:flow_diagram}). This plasma is
strongly overpressurized and expands on a timescale shorter than its
cooling time, thus driving a net outflow of multiphase gas with
outflow rates of a few $\times 10$ M$_{\odot}$ yr$^{-1}$
(\S\ref{sec:outflow}). In addition, the ablation of cloud material by
the outflowing hot medium may replenish the hot medium of the cocoon,
which would help maintaining a high pressure as the cocoon expands,
and enhance its lifetime. The energy cascade causes a momentum
transfer from large-scale bulk motion to turbulent motion on smaller
scales. Dynamical interactions between the different gas phases drive
fragmentation of the molecular clouds and turbulent motions between
and within individual fragments (second box in
Fig.~\ref{fig:flow_diagram}). This may lead to entrainment of parts of
the warm and cold medium in the hot wind as described for
starburst-driven winds \citep{heckman00}.

This interaction between the molecular gas and the outflowing plasma
creates a physical environment not very different from that produced
by the galaxy-wide shock in Stephan's Quintet \citep{guillard09,
  appleton06}.
Similar to the analysis of \citet{guillard09}, we postulate that
turbulent motion in the cocoon drives shocks with velocities that
depend on the gas density.  In a multiphase medium, differences in gas
densities between the cold and hot gas are up to factors of
$10^6$. For a given ram pressure, this translates into a factor $10^3$
in shock velocity.  The shocks maintain the amount of warm molecular
and ionized gas that are necessary to explain the observed luminous
line emission of the molecular and ionized gas (third box in
Fig.~\ref{fig:flow_diagram}). Due to the high density, shocks driven
into magnetized molecular gas will be slow, with velocities of a few
10s of km s$^{-1}$, so that H$_2$ will not be destroyed, but becomes a
main coolant. This is in agreement with our molecular diagnostic
diagram in Fig.~\ref{fig:moldiag} and with the pure rotational H$_2$
line ratios, which are consistent with excitation through slow shocks.

The gas cooling times for these shocks listed in
Table~\ref{tab:shock_mass_3C326} are very short, of order $10^4$
yrs. In order to maintain the gas at the observed temperatures, the
cocoon must inject energy over similar timescales. This is important,
since these timescales are significantly shorter than the free-fall
times of self-gravitating molecular clouds. This is evidence that the
gas cannot form gravitationally bound structures and stars, simply
because it is continually being stirred up by mechanical interactions
in the multiphase gas which are ultimately powered by the radio
source. We will discuss in \S\ref{ssec:dissipationtime} that the AGN
may maintain such conditions over significant timescales ($10^{7-8}$
yrs), and discuss astrophysical implications in \S\ref{sec:SF}.

\subsection{Efficiency of the energy transfer}
\label{ssec:canitbe?}

Simple energy conservation implies that our scenario for powering the
line emission and the outflow through mechanical interactions is only
realistic if the efficiency of the power transfer in each step is
$\le$ 1. We have estimates for the jet kinetic power ($\ge$few $\times
10^{44}$ erg s$^{-1}$, \S\ref{ssec:radiojet}) and the emission-line
luminosity of H$_2$ and HII ($\sim 10^{43}$ erg s$^{-1}$).  The
mechanical power necessary to drive the outflow is similar to the
emission-line luminosity at an order-of-magnitude level
(\S\ref{sec:outflow}).  Taken at face value, these estimates indicate
that at least $\sim$10\% of the jet kinetic luminosity is deposited
within the cocoon, which shows that our scenario of jet-powered line
emission is indeed energetically plausible. The efficiency of the
cocoon in driving an outflow -- the ratio between the mechanical
power carried by the outflow to the power thermalized in the cocoon --
may well be about 1, comparable to that derived from observations of
starburst-driven outflows \citep[e.g. M82][]{strickland09}. The
luminous line emission of 3C~326~N implies that a power comparable to
that of the bulk outflow is radiated away due to dissipation
 in the interstellar medium.

\subsection{Dissipation time and length of the H$_2$-luminous phase}
\label{ssec:dissipationtime}
If the radiation of luminous optical and infrared line emission is due
to the dissipation of the kinetic energy of the warm and cold
interstellar medium, then we can roughly estimate the dissipation time
of the kinetic energy of the gas, simply by relating the measured
kinetic energy and the emission-line luminosity. We will in the
following estimate this timescale, and compare with the characteristic
timescales of jet activity. 

In \S\ref{ssec:optdiagnostics} we found that the ionized gas, although
negligible in the mass budget, dominates the radiative energy budget
of the warm and cold interstellar medium with a bolometric luminosity
of $\sim 1\times10^{43}$ erg s$^{-1}$, whereas the molecular gas emits
a lower luminosity of $1\times 10^{42}$ erg s$^{-1}$. In
\S\ref{ssec:gaskinematics} we estimated that the kinetic energy of the
molecular gas is most likely $\sim 3 \times 10^{57}$ erg, about two
orders of magnitude more than that of the warm ionized gas (see also
Table~\ref{tab:summary}), and that it is not likely that this energy
budget will be dominated by rotation.

We will now constrain the range of plausible dissipation times,
assuming that the UV-optical emission of the ionized gas is powered by the
turbulent kinetic energy, and fully participates in the dissipation
process. In this case,

\begin{eqnarray}
\tau_{diss} =  \frac{3/2\ (M_{ion}\sigma_{ion}^2 + M_{mol}\sigma_{mol}^2)}
    {{\cal L}_{ion} + {\cal L}_{mol}} \sim 10^7 {\rm yr}
\end{eqnarray}

But the dissipation time could be significantly larger if the UV-optical lines
are not entirely powered by the turbulent energy. The molecular gas could be
in a region where the hot plasma has already been accelerated, so that the
kinetic energy of the hot wind would power the line emission through direct
interactions between the molecular gas and the surrounding warm and hot gas.
In this case, the bulk kinetic energy of the wind would be powering the line
emission, not the turbulence of the molecular gas. 
In the extreme case that the luminosity of the ionized gas is
fully dominated by the kinetic power from the outflow, it does not contribute
to the dissipation and the timescales simply follow from setting

\begin{eqnarray}
\tau_{diss} =  \frac{3/2\ (M_{mol}\sigma_{mol}^2)}{{\cal L}_{mol}} \sim 10^8{\rm yr}
\end{eqnarray}

\noindent (Note that we included He into our mass budget for both time
estimates.) These findings have two interesting consequences. First,
the dissipation time is much longer than the cooling time predicted by
our shock models for the molecular gas (\S\ref{sssec:warmmol}), which
is of order $10^4$ yrs. This may imply that the turbulent environment
of the cocoon feeds a mass and energy cycle similar to that described
by \citet{guillard09} for Stephan's Quintet, where the gas undergoes
many episodes of heating and cooling on ``microscopic'' scales,
effectively maintaining an equilibrium between different gas phases on
macroscopic scales. 

The large mass flow of a few 100 M$_{\odot}$ yr$^{-1}$ (which we
estimated in \S\ref{ssec:optdiagnostics} from the H$_{\beta}$
luminosity) is a natural outcome of this mass cycle.  It is too large
to be accounted for by the large-scale bow shock as the cocoon expands
through the galaxy, but may plausibly be produced by shocks that are
locally generated, as the molecular gas fragments move
relative to the low-density, hot medium. Shock velocities of order 250
km s$^{-1}$ inferred from the optical line ratios
(\S\ref{ssec:optdiagnostics}) are subsonic with respect to the hot
gas, but supersonic relative to the warm gas ($T< 10^6$ K).  Such a
cycling between warm and cold gas phases may also be important for the
exchange of momentum between the hot plasma and the H$_2$ 
gas and thereby for the entrainment of gas into the flow.
 
Second, a dissipation time of $\sim 10^{7-8}$ yrs is significant
compared to the lifetime of the radio jet (of order $10^7$ yrs,
\S\ref{ssec:radiojet}), or even the duty cycle of jet activity of
order $10^8$ yrs estimated from observations of 'rejuvenated' radio
sources \citep[e.g.,][]{schoenmakers00}.  If the dissipation time is
similar to the lifetime of the radio source, then the energy content
of the gas will reflect the {\it total energy injected by the radio
  source}, including the early stages of radio activity when the jet
was confined within the inner regions of the galaxy, and when the
interaction was likely to be particularly efficient. If it is even
longer, then it may influence the interstellar medium and
star-formation properties of the massive host galaxies of radio-loud
AGN even on timescales beyond their 'active' phase. This is 
possible if the gas luminosity drops at the end of the jet activity
period. In this case the dissipation timescales will be longer than
our estimates. Addressing this question in detail would require
observational constraints of the line emission in galaxies after their
radio-loud phase (radio relics) and is beyond the scope of this paper.

\section{Molecular gas and star formation}
\label{sec:SF}

(Cold) molecular gas and star formation are closely related: gas
surface densities and star-formation intensities (star-formation rate
surface densities) show a close relationship over several orders of
magnitude \citep['Schmidt-Kennicutt relation'][]{schmidt59,
  kennicutt98}. As we discussed previously, in H$_2$ luminous radio
galaxies, however, star-formation rates appear low compared to the
amount of available molecular gas. It is therefore interesting to
investigate whether these galaxies fall onto the same star-formation
law as 'ordinary' star-forming galaxies.

\subsection{Constructing a Schmidt-Kennicutt-like diagram from PAH and CO line emission}
The best way of estimating star-formation rates is, generally
speaking, the infrared continuum. However, in powerful radio galaxies
the 24$\mu$m and 70$\mu$m flux scales with AGN power
\citep{tadhunter07}, likely indicating that much of the dust is heated
by the AGN rather than star formation. We will therefore rather use
the PAH emission as an approximate tracer of star formation in our
galaxies. For the radio galaxies of \citet{ogle09} we have direct
measurements of the PAH fluxes from the IRS spectra, for star-forming
galaxies we will use the 8$\mu$m luminosity instead, which in these
galaxies is dominated by the 7.7$\mu$m PAH feature We use
  values corrected for the stellar contibution
  by \citet{calzetti07,roussel07} and have already outlined in \S\ref{ssec:moldiag} how we
  translate 8$\mu$m flux densities in PAH
  fluxes.

Although PAHs are also not very robust tracers of the star-formation
intensity generally, empirically they are found to be
equally robust as the 24$\mu$m flux when studying the integrated star formation rates in galaxies with high
metallicities. \citep{calzetti07}. The quality of our optical spectra
of 3C~326~N is not sufficient to safely determine a metallicity from
the absorption lines, but with its large stellar mass of order
$10^{11}$ M$_{\odot}$, 3C~326~N can safely be expected to have a high
metallicity. The same applies to H$_2$-luminous radio galaxies
generally.

In Fig.~\ref{fig:KSplot} we show a Schmidt-Kennicutt-like diagram
based on the PAH luminosity and CO surface brightness for 3C~326~N and
other H$_2$ luminous radio galaxies with published CO observations
\citep[][lists H$_2$ luminous radio galaxies with
  CO measurements in the literature]{ogle09}, as well as for star
forming and AGN host galaxies taken from SINGS \citep{roussel07}. For
galaxies where CO (and/or PAH) measurements are not spatially
resolved, we assume a radius of 2.5 kpc each. We are using the same
radius for PAH and CO emission, implying that molecular gas and PAH
emission originate from the same region of the galaxy, which is
astrophysically plausible.

Using equations (2) and (6) of \citet{calzetti07} we derive a
calibration of star-formation intensity as a function of 8$\mu$m
luminosity ($\nu L_{\nu}$) per $kpc^2$ or

\begin{equation}
\log{\Sigma_{SFR}} = -45.32 + 1.06\ \log{S_{8\mu m}}
\end{equation}
where $\Sigma_{SFR}$ is the star-formation intensity in M$_{\odot}$
yr$^{-1}$ kpc$^{-2}$ and S$_{8\mu m}$ the luminosity at 8$\mu$m in erg
s$^{-1}$ kpc$^{-2}$ corrected for the stellar continuum
\citep[see][for details]{calzetti07}. We also use this relationship to
re-calibrate the star-formation law of \citet{kennicutt98} in terms of
PAH luminosity, finding

\begin{equation}
\log S_{8\mu m} = 31.44 + 1.32 \ \log{\Sigma_{gas}}
\end{equation}

where $\Sigma_{gas}$ is the molecular gas
surface density in M$_{\odot}$ kpc$^{-2}$. This relationship is somewhat
less steep than the original \citeauthor{kennicutt98} law (which has a
power-law index 1.4), simply because PAH surface brightness and
star-formation intensity do not scale linearly. 

\subsection{Is the star-formation efficiency low in H$_2$ luminous radio galaxies?} 
The Schmidt-Kennicutt law is shown as the black solid line in
Fig.~\ref{fig:KSplot}, and is a good representation of the SINGS
galaxies with CO emission line observations \citep[][]{roussel07}. In
strong contrast, 3C~326~N as well as the other H$_2$ luminous radio
galaxies shows a pronounced offset by roughly 1-1.7 dex (a
factor 10-50) towards lower PAH fluxes, but with a very similar
slope. If the 'standard' CO conversion factor roughly applies, and if
PAH emission traces star formation in a similar way as in 'ordinary'
star-forming galaxies, then our result implies that the star formation
 in H$_2$ luminous radio galaxies is about a factor
10-50$\times$ less efficient than in 'ordinary' star-forming
galaxies. But even if these assumptions were not strictly appropriate,
the basic result would persist, that the conditions of the
interstellar medium in 3C~326~N and other H$_2$ luminous radio galaxies
are significantly different from those in galaxies with 'ordinary'
star formation properties.

\subsubsection{Caveats}
Fig.~\ref{fig:KSplot} is based on several assumptions, each of which
introduces systematic uncertainties. This is why we prefer not to state a
specific offset but rather give a range of possible offsets from the
Schmidt-Kennicutt law. We will now discuss these uncertainties.

One obvious caveat in our conclusion is that we only used the CO
measurements to estimate a gas mass, neglecting the warm
H$_2$. Including the warm molecular gas in our mass estimate would
only increase the offset of the H$_2$ luminous galaxies relative to
the SINGS sample, for example by 0.3 dex for 3C~326~N. Changes in the
CO-to-H$_2$ conversion factor would have to be as large as the offset
to move the H$_2$ luminous galaxies onto the nearby relationship, and
would again not change the basic conclusion, namely that the physical
gas conditions in these galaxies are markedly different from those in
'ordinary' star-forming galaxies. In this case, the CO-to-H$_2$
conversion factor would have to be {\it lower} by a factor 30 than in
the SINGS galaxies, and a factor 6 lower than in ULIRGs \citep{downes98}.

Another possible caveat is that we adopted a fiducial radius of 2.5
kpc for the galaxies where we only have integrated measurements (empty
red triangles in Figure~\ref{fig:KSplot}). Varying this radius by
large amounts (greater than factors 2-3) will mildly affect the
position of individual galaxies, but is not sufficient to put them
onto the 'ordinary' Schmidt-Kennicutt ridge. This limitation can
easily be overcome by collecting larger samples with high-resolution,
high-quality CO observations. Galaxies shown as red filled
triangles have spatially-resolved CO measurements and are not affected
by this effect. We adopted the same radii for PAH and CO emission,
which corresponds to the astrophysical assumption that both are
related to the same star-forming regions. 

We may also wonder whether this offset could reflect a net deficit of
PAH emission? We will rely on the close analogy with the
H$_2$-luminous shock in Stephan's Quintet to address this concern. For
Stephan's Quintet, \citet{guillard10} show that the PAH and mid-IR
continuum associated with the warm H$_2$ emission is consistent with
the expected emission from the warm molecular gas, assuming that the
dust-to-gas mass ratio, and the fraction in PAHs and very small grains
have roughly the same values as in the Milky Way. 

The ratios between H$_2$, CO and PAH
luminosities in 3C~326 are very similar to those measured in Stefan's
Quintet (Figure~\ref{fig:moldiag}), so that similar arguments may
apply to 3C~326~N.  For a standard (Galactic) dust mass fraction in
PAHs, the weakness of the PAH emission implies a UV field comparable
to that observed in the intergalactic shock in Stefan's Quintet. 
The UV intensity in Stephan's Quintet, 
measured with GALEX, is not very different from that in the Solar
Neighborhood (1.4 in Habing units\footnote{The Habing field is equal
  to $2.3~{\rm erg \, s^{-1} \, cm^{-2}}$ at $\lambda = 1530~\AA$
  \citep{habing68}}).  
We
used Starburst99 \citep{leitherer99} to deduce an upper limit on the
UV radiation field from the measured upper limit on the star-formation
rate \citep[$0.07~{\rm M_\odot \, yr^{-1}}$, inferred by][from the PAH
  and mid-infrared dust continuum of 3C~326~N]{ogle07}.  Assuming that
the molecular gas is within a sphere of 2.5~kpc radius (the approximate
extent of the CO emission), we find an upper limit on the UV radiation
field of about 1 Habing unit. Thus, the weakness of the PAH and dust mid-IR
emission is consistent with the low star-formation rate, and does not
require the molecular gas to be PAH poor. Measuring the dust mass and
temperature directly from the far-infrared luminosity is now possible
with Herschel. This will allow for more robust estimates of the
star-formation rates, and give direct constraints on the CO-to-H$_2$
conversion factor. These measurements will be critical to better
estimate the star-formation efficiencies in H$_2$ luminous galaxies.

\subsubsection{Implications}
We are not the first to note that star formation in radio galaxies
with CO detections appears suppressed relative to radio-quiet
galaxies. \citet{okuda05} and \citet{papadopoulos07} found the same
effect for 3C31 and 3C293, respectively. \citet{koda05} identify
star-forming knots embedded in the molecular disk of 3C31, pointing
out that, although the disk appears globally stable against gas
collapse and star formation, this is not sufficient to quench all star
formation.  The position of 3C31 in Fig.~\ref{fig:KSplot} suggests
that, nonetheless, the overall star formation efficiency in this
galaxy is considerably lower than in 'ordinary' star-forming galaxies.

Our previous discussion of how the radio source influences the energy
budget of the interstellar medium of 3C~326~N (\S\ref{sec:enmom})
suggests that heating through the radio source may be an attractive
non-gravitational mechanism, at least in galaxies that host radio-loud
AGN, which may enhance the turbulence and heating in a molecular disk,
preventing or suppressing gravitational fragmentation and gas collapse
and ultimately star formation.
In Cen~A (NGC 5128) \citet{neumayer07} find that the ro-vibrational
H$_2$ emission lines are broader than expected from the stellar
velocity dispersion, although they have an overall similar velocity
field. This would certainly agree with a scenario where a small
fraction of the energy injected by the AGN into the multiphase
interstellar medium of the host galaxy will finally contribute to
increasing the turbulence in the molecular disk. Detailed follow-up
observations of a sufficiently large sample of H$_2$-luminous galaxies
are certainly necessary to substantiate this speculation.

We did not find evidence for positive AGN feedback enhancing the star
formation efficiency relative to 'ordinary' star forming galaxies, as
proposed by, e.g., \citet{vanbreugel85, begelman89, mellema02,
fragile04, croft06, silk09}.  \citet{vanbreugel85, schiminovich94,
croft06, elbaz09} claim evidence for star formation associated with
radio jets outside the host galaxy in a few individual nearby systems,
which may be triggered by the jet. Indeed, turbulence can trigger star
formation locally where gravitationally-bound giant molecular clouds
are otherwise not able to form \citep{klessen00}. If positive feedback
dominated in our galaxies, we would expect an offset to higher
star-formation intensities for a given gas surface density in
Fig.~\ref{fig:KSplot}, which is not the case, as all radio galaxies are shifted towards lower star-formation efficiencies.

Our overall results suggest that the suppression or enhancement of
star formation depends critically on the details of the
'micro-'physics of the gas on small scales, which is beyond the
capabilities of current numerical models and requires detailed
observations of the multiphase gas in AGN host galaxies, which are
becoming possible only now with Spitzer, Herschel, ALMA, and JWST. In
particular, it is unclear whether our results can be easily
extrapolated to (radio) galaxies at high redshift, which have copious
amounts of molecular gas traced through CO line emission
\citep[e.g.,][]{papadopoulos00, debreuck03, debreuck05, nesvadba09},
but whose gas conditions are likely very different from those in
3C~326~N. If the kinetic energy in these galaxies can be dissipated
rapidly enough to allow for gas collapse and star formation, then
feedback may be positive.

\section{Implications of this scenario for galaxy evolution}
\label{sec:implications}
The energy injected by powerful AGN into the interstellar medium and halo of
galaxies may play an important role in determining the characteristics of
galaxies in the local Universe.  For example, the high metallicities,
luminosity-weighted stellar ages, and relative abundances of $\alpha$ elements
relative to iron suggest rapid, and truncated star formation in the early
Universe followed by a long phase of (mostly) passive evolution
\citep[e.g.,][]{pipino04} for massive galaxies. These observations
are consistent with a phase of powerful AGN driven outflows in the early
Universe. In fact, studies of powerful radio galaxies at z$\sim 2$ with
rest-frame optical integral-field spectroscopy reveal energetic outflows which
have the potential of removing a significant fraction of the interstellar
medium of a massive, gas-rich galaxy within the short timescales necessary to
explain the super-solar [$\alpha$/Fe] ratios and old luminosity weighted ages
observed in massive ellipticals \citep{nesvadba06,nesvadba07,nesvadba08}. This
``quenching'' of star formation by powerful radio jets is in broad agreement
with theoretical models explaining the characteristics of massive
galaxies.  However, in addition to this quenching phase, the
radio-loud AGN may also assist in maintaining low star-formation rates
over cosmological timescales by inhibiting subsequent gas cooling --
``maintenance phase''. This last phase is important since, even in the
absence of mergers, subsequent gas infall and return from the evolving
stellar population will replenish the reservoir of cold gas which must
be prevented (at least partially) from forming stars.

In the scenario we presented above, roughly equal amounts of energy
are being dissipated through turbulence, and are powering an outflow
of warm neutral and ionized gas. Gas that does not escape from the
halo of the host galaxy, will likely cool and rain back onto the
galaxy.  For cool-core clusters, \citet{revaz08,salome08} propose that
molecular filaments may represent gas that has been lifted from the
galaxy, has cooled, and may be ``raining back down'' into the deepest
part of the gravitational potential. A similar mechanism may apply to
individual galaxies like 3C~326~N. In addition, the stellar population
will continue to feed the interstellar medium with material which will
also cool and descend into the potential well, if the material will
not have been ablated previously. This may once again fuel an AGN,
leading to a self-regulating AGN fueling cycle, even in the absence of
galaxy merging.

We have so far focused our discussion on 3C~326~N which is the
most extreme H$_2$ luminous radio galaxy known and therefore well suited to
study the physical mechanism responsible for the H$_2$ line emission. However,
galaxies are complex astrophysical objects, where a wide spectrum of physical
and astrophysical phenomena may act in parallel. In order to illustrate how
different formation histories may lead to different observational signatures
within a common physical framework, we will in the following contrast 3C~326~N
with 3C293 at z=0.045. We have chosen 3C293 because it has particularly
luminous CO line emission \citep[][ and unlike 3C~326]{evans99} and with CO
line ratios suggesting the molecular gas is highly excited
\citep{papadopoulos08}. It also has a significant outflow of neutral gas
\citep{morganti03,morganti05,emonts05}. Its H$_2$ emission-line luminosity,
warm molecular gas mass and optical line ratios are very similar to those of
3C~326~N \citep{ogle09}, which suggests that 3C293 may be a 'gas-rich analog'
and thus a foil of 3C~326~N.

\subsection{Differing Star-formation Histories: Clues to the importance
of molecular gas content?}
\label{ssec:stpops}

\begin{figure}
\centering
\includegraphics[width=0.7\textwidth]{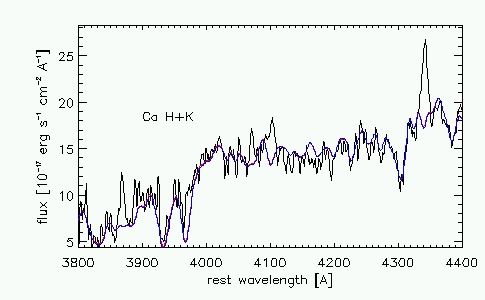}\\
\includegraphics[width=0.7\textwidth]{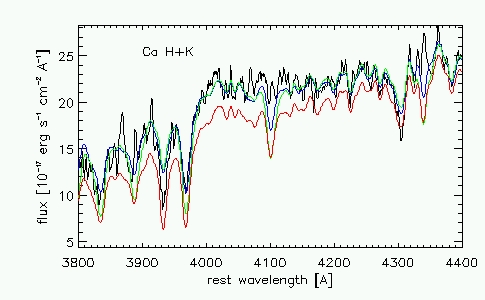}\\
\caption{Wavelengths near the 4000 \AA\ break are most sensitive to the
star-formation history of 3C~326~N {\it (upper panel)} and 3C293 {\it (lower
 panel)}. The black line marks the SDSS spectrum in each panel. Red, blue,
and green lines mark the \emph{Starlight} spectral fits for different
star-formation histories, see \S\ref{ssec:stpops} for details. {\it Upper
  Panel:} Red -- Stellar population older than $10^{10}$ yr. Blue -- Old and
intermediate-age (few $10^9$ yrs) stellar population. Both scenarios fit the
data equally well. {\it Lower Panel:} Red -- Stellar population older than
$10^{10}$ yr. Green -- Old  and intermediate-age (few
$10^{8-9}$ yr) population. Blue -- Old , intermediate-age
, and young ($10^7$ yr) stellar population. A single
star-formation episode $10^{10}$ yrs ago does not appear as a good fit to
the data; adding a young stellar population (consistent with the observed
starburst) improves the fit to the Ca
H$+$K lines, although significant differences remain, especially for the blue
component.} \label{fig:stpops} 
\end{figure}

\begin{figure}
\centering
\includegraphics[width=0.9\textwidth]{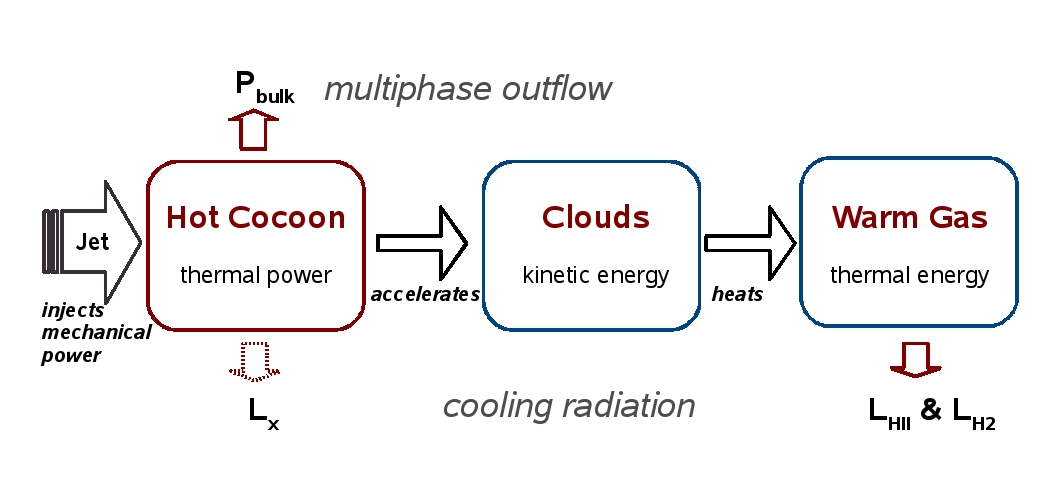}\\
\caption{
This diagram outlines the energy flow within the hot ``cocoon'' that we
propose to account for the HII and H$_2$ line emission and the wind. The
flow is powered by the interaction of the jet with the multiphase interstellar
medium of the host galaxy (arrow to the left).  It feeds three energy
reservoirs: {\it (boxes from left to right:)} the thermal energy of the hot
plasma surrounding the relativistic radio source, the turbulent kinetic energy
of the clouds embedded in the plasma and the thermal energy of the warm HII
and H$_2$ gas.  The arrows at the bottom represent the ``cocoon'' energy loss
by radiative cooling, that to the top the transformation of thermal energy
into bulk kinetic energy of the outflowing gas.  The dissipation of the
turbulent kinetic energy powers the H$_2$ and HII line emission. The high
pressure plasma expands on a time scale shorter than its cooling time through
X-ray emission driving a multiphase outflow. Our observational results suggest
that the kinetic power of the outflow and the power radiated by the HII and H$_2$
gas are both $\rm \sim 1 \times 10^{43} erg\ s^{-1}$. The power input
represents $\sim 10 \%$ of the mechanical power of the jet.} 
 \label{fig:flow_diagram}
\end{figure}

\begin{figure}
\centering \includegraphics[width=0.7\textwidth]{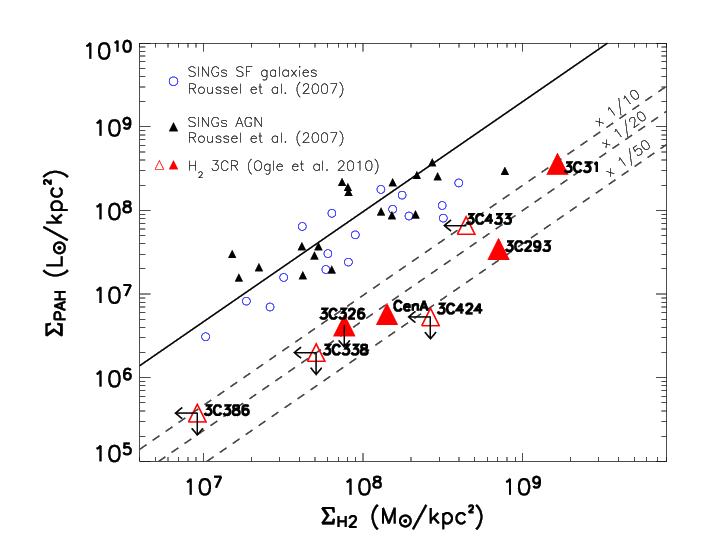}\\
\caption{PAH intensity as a function of molecular gas mass surface
  density measured from the CO(1-0) emission-line intensity.  The
  black solid line shows the Schmidt-Kennicutt law
  \citep[][]{kennicutt98}, where the star-formation intensity is
  translated into a PAH intensity using the results of
  \citet{calzetti07}. SINGS star-forming and AGN host galaxies from
  \citet{roussel07} are shown as small black triangles and small blue
  circles, respectively, and fall onto the relationship. In stark
  contrast, 3C~326~N and other H$_2$ luminous radio galaxies
  \citep[big red triangles; PAH fluxes are taken from][]{ogle09} have
  a marked offset by roughly factors 10-50 towards lower PAH
  intensities for a given molecular gas mass (black hatched lines),
  but are consistent with having a similar slope as the 'standard'
  Schmidt-Kennicutt law. This offset would become larger if we had
  included the warm molecular gas (e.g., by 0.3 dex for
  3C~326~N). (Filled red triangles mark galaxies with spatially
  resolved CO detections, empty triangles mark galaxies with
  integrated measurements. Arrows mark galaxies with sensitive upper
  limits.)}
 \label{fig:KSplot} 
\end{figure}

3C~326~N and 3C293 have very different star-formation histories. The spectrum
of 3C~326~N is consistent with an old stellar population with an age of
$>10^{10}$ yrs. At most a
small fraction of the total stellar mass, $\le 5$\%, formed recently ($\sim
1-2\times 10^9$ yrs; Fig.~\ref{fig:stpops}). Within the 3\arcsec\ aperture of
the SDSS fiber, this would correspond to $3\times 10^9 M_{\odot}$ of stars
formed during the last 1-2 Gyrs, implying a star-formation rate of about $1
M_{\odot}$ yr$^{-1}$ for a constant star-formation history. We note that this
is more than a factor 10 greater than the upper limit on the current
star-formation rate of $\le$0.07 M$_{\odot}$ yr$^{-1}$ estimated from the
infrared emission.

For 3C293 we find a significantly more complex star-formation history
\citep[see also][]{tadhunter05}.  Namely $\sim 80$\% of the stellar mass of
3C293 was formed at high redshift, $\sim 20$\% were formed in a more recent
star-formation episode about $1-2\times 10^9$ yrs ago \citep[][ found similar
  results]{tadhunter05}.  In addition, the equivalent widths and ratio of the
Ca H$+$K absorption lines require a very young population with an age of $\sim
10^7$ yrs, consistent with the starburst implied by far-infrared luminosity
\citep{papadopoulos08}.  The current burst contributes $<$1\% to the total
mass of 3C293, and the star-formation rate of SFR$^{293}_{popsyn}\sim$4
M$_{\odot}$ yr$^{-1}$ is broadly consistent with the current SFR$^{293}_{IR}$=
7 M$_{\odot}$ yr$^{-1}$ estimated from the infrared luminosity of 3C293
\citep{papadopoulos08}.  To evaluate whether this fit is unique, we tested
alternative star-formation histories with only one or two of the three
populations. Fig.~\ref{fig:stpops} illustrates that the three-component fit
indeed is a significantly better representation of the data.  The population
synthesis fits also yield stellar mass estimates, but we need to correct for
the 3\arcsec\ fibre size of the SDSS, which is much smaller than the size of
the galaxies.  Doing this, we find stellar masses of M$_{stellar}^{326N} =
3\times 10^{11}$ M$_{\odot}$ and M$_{stellar}^{293}=7\times
10^{11}$M$_{\odot}$ for 3C~326~N and 3C293, respectively.

\subsection{Origin of the gas}

3C293 has a $\sim 10\times$ larger reservoir of gas traced through CO
line emission than 3C~326~N. What is the likely origin of the gas in
these galaxies? Both galaxies have nearby galaxies with which they are
perhaps interacting.  For the pair 3C~326~N/S, we find a very low gas mass
and red optical colors for both galaxies, suggesting this is a gas-poor
system.  Indeed our estimate of the stellar and gas mass suggests a
gas fraction of only $\sim$ 1\%. The star-formation history of 3C~326~N
appears quiescent and does not indicate that the galaxy was undergoing a
starburst due to the interaction.  Nonetheless, we may expect 
the accumulation of a significant amount of cold gas due to mass loss from
its stellar population ($\sim$1 M$_{\sun}$ yr$^{-1}$) as well as perhaps
accretion from the surrounding halo. How much gas will this likely be?

We use the analysis of \citet{kennicutt94} to estimate the amount of
gas returned by a single-age stellar population into the interstellar
medium over 10 Gyr of passive evolution.  \citet{kennicutt94} find that
depending on the details of the star-formation history and initial
mass function, as much as 30-50\% of the stellar mass formed in the
burst may be returned into the ISM during 10 Gyr.  From their Fig.~8 we
estimate that the gas return after the first 0.5-1 Gyrs will be of order
a few percent of the stellar mass. For the stellar mass of 3C~326~N, and
ignoring any other mechanism like subsequent star formation or feedback,
this would correspond to a total gas mass of several 10$^9$ M$_{\odot}$
up to$\sim 1\times 10^{10}$ M$_{\odot}$, factors of a few more than the
amount of cold and warm gas we observe in 3C~326~N. In addition, with
an infall rate of $\sim$ 0.1-1 M$_{\odot}$ yr$^{-1}$ of gas accreted
from the halo \citep[which appears plausible for early and late-type
galaxies; ][]{sancisi08} over $10^{10}$ yrs, this gas mass may roughly
double. An infall rate of $\sim 0.1 M_{\odot}$ yr$^{-1}$ of neutral gas has
also been estimated by \citet{morganti09} from the observation
of a narrow, redshifted HI cloud seen in absorption and emission in the
nearby, radio-loud early-type galaxy NGC315. 

Hence, each of these processes alone could easily explain the observed
gas mass of 3C~326~N. For 3C293 however, similar estimates of stellar
mass loss and gas accretion would be inefficient to explain the amount
of gas necessary for forming the $\sim 10^{11}$ M$_{\odot}$ of stellar
mass during the last Gyr, which is certainly consistent with
the assumption of a merger-triggered starburst for this galaxy. So the
difference could be in the ``mass accumulation histories'' of these
two AGN.

\subsection{Maintaining the low fraction of recent star-formation}
\label{ssec:maintenance}
Despite these plausibly different accretion histories, both of the galaxies
apparently have low star-formation efficiencies.  We estimated that a total of
$1-2 \times 10^{10}$ M$_{\odot}$ of warm/cold gas would accumulate in a
massive early-type galaxy such as 3C~326~N over a Hubble time while 3C293
acquired its gas during an interaction/merger.  Is it plausible that the gas in
both sources will be heated and substantial fractions removed by the
mechanical energy of the jets?

\citet{best05} find that the number of radio sources is a strong function
of radio power and stellar mass of the host galaxy for a sample of
2000 early-type galaxies with FIRST and NVSS observations taken from the
SDSS. For example, for galaxies with stellar mass of $\sim 3\times10^{11}$
M$_{\odot}$ they estimate that $\sim 10$\% of all galaxies have radio
powers $\ge 10^{24}$ W Hz$^{-1}$, and propose that this may represent
a duty cycle, where any given galaxy of this stellar mass will host a
similarly powerful radio source for 10\% of the time.

Statistically speaking, over 10 Gyrs, a given AGN in a few $\times 10^{11}$
M$_{\odot}$ galaxy would be active for a total of $\sim 10^9$ yrs (this total
'activity period' may consist of many radio-loud phases with
intermediate, quiescent phases). With the same
reasoning as in \S\ref{ssec:radiojet} we estimate that a power $\ge 10^{24}$ W
Hz$^{-1}$ measured at 1.4 GHz will correspond to a kinetic power of order
$10^{43}$ erg s$^{-1}$, equivalent to a total energy of a few $\times 10^{59}$ erg for a
total activity time of $10^9$ yrs (about 10\% of a Hubble time). A few $\times
10^{58}$ erg of energy are necessary to unbind $10^{10}$ M$_{\odot}$ in gas
from the potential of a galaxy with few $\times 10^{11}$ M$_{\odot}$ in
stellar mass \citep{morganti05,nesvadba06,nesvadba08}. Based on these simple
energy considerations, it appears overall plausible that AGN in massive
galaxies may regulate gas cooling through the energy injected by their radio
sources over a Hubble time as expected for the ``maintenance" mode.

However, in addition to the energy requirement, we also have a
timescale requirement, since gas cooling must be inhibited inbetween
phases of radio-loud AGN activity, if AGN are to shape the properties
of the ensemble of (massive) galaxies. Several arguments may suggest
that this could indeed be possible. First, the dissipation timescales
we found in \S\ref{ssec:dissipationtime} of order $10^{7-8}$ yrs imply
that significant fractions of the molecular gas will remain warm
during all of the lifetime of the radio jet, and possibly also for a
significant time afterwards (if the upper estimate, albeit not very
precise, is more appropriate). Interestingly, studies of 'rejuvenated'
jet activity suggest timescales of about $10^8$ years between activity
episodes \citep{schoenmakers00}, similar to the upper range of the
H$_2$ dissipation time. Second, the mechanical energy injected by
radio-loud AGN is sufficient to accelerate and perhaps remove
significant amounts of gas. Even if parts of the entrained material
will not reach escape velocity, its density will decrease due to the
larger volume. The mean free path of particles between collisions will
be greatly increased, making the dissipation timescales accordingly
longer and delaying the time until this gas will cool and rain back
onto the galaxy. Third, the expelled gas may be reheated by mixing
with the hot halo gas. Fourth, the molecular gas we observe may only
be transitory, where the molecules are formed by the ram pressure of
diffuse gas \citep{glover07} as a result of the dissipation of the
turbulent energy. This process does not necessarily lead to the
formation of gravitationally bound clouds, in which case the H$_2$
destruction through photodissotiation may dominate over the H$_2$
formation, when the turbulent and thermal pressure in the cocoon drop.

\section{Summary}
We presented a detailed analysis of the physical gas conditions in the
nearby powerful, H$_2$-luminous radio galaxy 3C~326~N, which does not
show the signatures of active star formation in spite of few $\times
10^9$ M$_{\odot}$ in molecular gas. Our main goals
were to investigate the gas conditions in this galaxy, to unravel the
astrophysical mechanism which powers the remarkably bright
mid-infrared H$_2$ emission, and to elucidate why this galaxy is not
forming stars in spite of a considerable reservoir of molecular
gas. To this end, we combined newly obtained CO(1-0) millimeter
spectroscopy of 3C~326~N obtained at the Plateau de Bure Interferometer
with the Spitzer-IRS spectroscopy by \citet{ogle07,ogle09} and optical
spectra from the SDSS. Our main conclusions are as follows: 

\noindent (1) We compare gas masses for warm and cold gas, estimated
from Spitzer mid-infrared spectroscopy and IRAM millimeter imaging
spectroscopy. We find that most of the molecular gas in 3C~326~N is
warm ($\sim 2\times 10^9$ M$_{\odot}$, compared to $1.5\times 10^{9}$
M$_{\odot}$ estimated from our CO(1-0) observations for a ''standard''
CO-to-H$_2$ conversion factor). This ratio of warm to cold gas mass is
about 1 to 2 orders of magnitude larger than that found in
star-forming galaxies, indicating that the gas is in a distinct
physical state, which may account for the low star-formation
efficiency.

\noindent (2) We introduce a new 'molecular' diagnostic diagram based
on the pure-rotational H$_2$, CO and PAH emission to show that most of
the line emission from molecular gas in 3C~326~N is not produced by UV
heating. We argue that the gas is likely to be powered by the dissipation
of mechanical energy through shocks. The same is found for the ionized
gas.

\noindent (3) Interstellar Na~D absorption marks an outflow of neutral
gas at velocities of up $\sim -1800$ km s$^{-1}$, and most likely mass
and energy loss rates of 30$-$40 M$_{\odot}$ yr$^{-1}$ and $\sim
10^{43}$ erg s$^{-1}$, respectively. These values are similar to those
found in starburst-driven winds, but this is the first time that such
an outflow is detected in the Na~D line of a galaxy which does not
have strong star formation. The star-formation intensity in 3C~326~N
is orders of magnitudes below what is necessary to drive a
wind. Similarly, the line profile cannot be explained through the
interaction between 3C~326~N and 3C~326~S.

\noindent (4) Based on these observations we propose a scenario where
the outflow and H$_2$ line emission are intricately related. It
represents an extension of the well-explored 'cocoon'-model of
interactions between radio jets and the ambient gas, and includes the
physics of the molecular gas. In this scenario,the H$_2$ line emission
is powered by turbulence induced within dense clouds that are embedded
in the expanding cocoon, with a dissipation time of $\sim 10^{7-8}$
yrs. Dissipation times of $10^8$ yrs are in rough agreement with the
duty cycle of jet activity estimated from rejuvenated radio sources.

\noindent (5) Comparing PAH and CO surface brightness (in analogy to
the ``Schmidt-Kennicutt'' diagram), we find a significant offset
between 3C~326~N and other H$_2$-luminous galaxies towards lower PAH
surface brightnesses. This may suggest that star-formation
efficiencies in these galaxies are lower by roughly a factor $\sim
10-50$ than those in 'ordinary' star-forming galaxies.

\noindent (6) Generalizing our results for 3C~326~N we find that
outflows during similar radio-loud episodes in massive galaxies may
balance the secular supply in cold gas through accretion and the mass
return from evolved stars over a Hubble time.  If radio-activity is a
common, but episodic property of most massive early-type galaxies as
suggested by \citet{best06}, then the long dissipation times of up to
$\sim 10^8$ yrs may have consequences for the population of massive
galaxies as a whole.

We emphasize that this analysis can only be the first step towards an
understanding of how radio-loud AGN regulate the physical conditions
of the molecular gas and ultimately star formation in massive
galaxies. It illustrates the importance of following a
multi-wavelength approach if we seek to relate the physics and
astrophysics of AGN feedback with the physical conditions of the
interstellar medium including molecular gas in the host galaxy and the
regulation of star formation.

\begin{acknowledgements}
We would like to thank the staff at IRAM for carrying out the
observations. We are particularly grateful to the referee, C. De
Breuck, whose comments helped significantly improve the paper, and to
Luc Binette and Geoff Bicknell for helpful discussions. This work was
supported by the Centre National d'Etudes Spatiales (CNES). NPHN also
acknowledges financial support through a fellowship of the Centre
National d'Etudes Spatiales (CNES). IRAM is funded by the Centre
National de Recherche Scientifique, the Max-Planck Gesellschaft and
the Instituto Geografico Nacional. This work is partly based on
observations made with the Spitzer Space Telescope, which is operated
by the Jet Propulsion Laboratory, Caltech, under NASA contract 1407.
\end{acknowledgements}

\bibliographystyle{aa}
\bibliography{nesvadba2010_326}

\clearpage
\onecolumn

\clearpage
\begin{table} 
\begin{center} 
\begin{tabular}{lccc} 
\hline 
\hline 
Aperture    & $\nu_{obs}$ & Peak   &  Width \\ 
         & [GHz]     & [mJy]  &   [km/s] \\ 
\hline 
2.5\arcsec$\times$2.1\arcsec\ & 105.744 & 1.1$\pm$0.2&351$\pm$98\\
5\arcsec$\times$5\arcsec\     & 105.744 & 1.5$\pm$0.3&351$\pm$98\\
\hline 
\end{tabular} 
\caption{Observational results: CO(1--0) line-fit parameters of (1) a
  spectrum integrated over a region of 2.5$''$ by 2.1$''$ and (2), a
  spectrum integrated over a 5\arcsec$\times$5\arcsec\ aperture
  corresponding to the slit width of the IRS spectrum (dashed line in
  Fig. \ref{fig:CO_spectrum})}
\label{tab:co1} 
\end{center} 
\end{table} 

\begin{table} 
\begin{center} 
\begin{tabular}{lcccc} 
\hline 
\hline 
Aperture     & redshift & I$_{line}$          & L$^\prime$                    & F$_{\rm{cont}}$ \\ 
             &          & [Jy\,km\,s$^{-1}$]  &  [K\,km\,s$^{-1}$\,pc$^{2}$]  &  [mJy] \\ 
\hline 
2.5\arcsec$\times$2.1\arcsec & 0.0901 & 0.5$\pm$0.1& 2$\pm$1$\times$10$^{8}$  &   0.99\\ 
5\arcsec$\times$5\arcsec     & 0.0901 & 1.0$\pm$0.2& 3.8$\pm$1$\times$10$^{8}$  &   0.99\\ 
\hline 
\end{tabular} 
\caption{Observational results: Intensity of the CO(1--0) emission line after
  subtraction of a point-like continumm source with a fixed value of 0.99 mJy
  at the position of the cm radio-source (Rawlings et
  al. 1990). L$^{\prime}_{CO}$ was computed with the formula of Solomon et
  al. (1997). Results are given for (1) a spectrum integrated over a region of
  2.5$''$ by 2.1$''$  and (2) a spectrum integrated over a
5\arcsec$\times$5\arcsec\ aperture corresponding to the slit width of the IRS
spectrum (dashed line in Fig.~\ref{fig:CO_spectrum}).}
\label{tab:co2} 
\end{center} 
\end{table} 

\begin{table}
\centering
\begin{tabular}{ccccc}
\hline
line & $\lambda_{rest}$ &  redshift &FWHM & flux \\
(1)  & (2)             & (3)             & (4)       & (5) \\
\hline
[OII]      & 3727.00 & 0.09019 & 623 & 9.62 \\
H$\beta$   & 4861.30 & 0.09017 & 610 & 2.44 \\
$[$OIII$]$ & 4959.90 & 0.08991 & 609 & 0.58 \\
$[$OIII$]$ & 5006.90 & 0.09011 & 609 & 1.77 \\
$[$OI$]$   & 6300.30 & 0.09013 & 612 & 2.22 \\
$[$OI$]$   & 6363.80 & 0.09013 & 612 & 0.53 \\
$[$NII$]$  & 6548.10 & 0.09014 & 601 & 2.38 \\
H$\alpha$  & 6563.30 & 0.09007 & 596 & 9.30 \\
$[$NII$]$  & 6584.10 & 0.09003 & 601 & 7.17 \\
$[$SII$]$  & 6716.40 & 0.09019 & 605 & 4.86 \\
$[$SII$]$  & 6730.80 & 0.09019 & 606 & 3.68 \\
\hline
\end{tabular}
\caption{Emission-line fluxes in 3C~326~N. Column (1) -- Line
  ID. Column (2) -- Rest-frame wavelength in \AA. Column (3) --
  Redshift. Column (4) -- Full width at half maximum in km s$^{-1}$,
  corrected for an instrumental resolution of 167 km s$^{-1}$,
  corresponding to a spectral resolving power of R=1800. Uncertainties
  are a few 10s of km s$^{-1}$. Column (5) -- Integrated line flux in
  units of 10$^{-15}$ erg s$^{-1}$ cm$^{-2}$. All values are derived
  after subtracting the best-fit stellar continuum (see text for
  details). Uncertainties are a few \% for all lines. For
  [OII]$\lambda\lambda$3727, additional uncertainties are due to the
  unknown line ratio between the two components of the doublets. These
  are insignificant for the FWHM and flux estimate, and amount to
  $\sim 70$ km s$^{-1}$ for the redshift, which however is not one of
  the important quantities for our analysis.}\label{tab:3c326emlines}
\end{table}

\begin{table}
\centering
\begin{tabular}{lccc}
\hline 
Line ratio & observed & shock & shock \& precursor \\ 
(1)  & (2) & (3) & (4) \\
\hline
H$\beta$/H$\alpha$ & 0.26 & 0.33 & 0.33 \\ 
$[$OIII$\,]$/H$\beta$ & 0.72 & 0.82 & 1.48 \\ 
$[$OI$\,]$/H$\alpha$ & 0.24 & 0.22 & 0.13\\ 
$[$NII$\,]\lambda$6584/H$\alpha$ & 0.77 & 0.68 & 0.47\\ 
$[$SII$\,]$/H$\alpha$ & 0.61 & 0.52 & 0.32 \\
$[$SII$\,]\lambda$$\lambda$6716,6731 & 1.32 & 1.23 & 1.24\\ 
$[$NeII$\,]$(12.81$\mu$m)/H$\alpha$ & 0.46 & 0.45 & 0.28 \\ 
$[$NeII$\,]$(12.81$\mu$m)/[NeIII](15.6$\mu$m) & $>$1.8 & 2.2 & 0.73 \\
\hline
\end{tabular}
\caption{Measured emission-line ratios in 3C~326~N and expected ratios
  for a pure-shock model and a model assuming a shock and precursor
  \citep{allen08}. The line ratios are consistent with shock
  velocities of v$=$250 km s$^{-1}$, a pre-shock density of 1
  cm$^{-1}$, and a magnetic parameter of B/$\sqrt(n)\le$ 1 $\mu$G
  cm$^{-3/2}$. Column (1) -- Line ID. Column (2) -- Observed line
  ratio. Column (3) -- Pure shock model. Column (4) -- Shock and
  precursor. $[$NeII$]$ and $[$NeIII$]$ fluxes are taken from
  \citet{ogle09}.}
\label{tab:optshock}
\end{table}

\begin{table}
\centering
\begin{tabular}{lcccc}
\hline
                &  Molecular Gas & Ionized Gas & Radio source & AGN X-ray \\
                & (1) & (2) & (3) & (4) \\
\hline
obs. luminosity & 41.9 & 41.3 & 44.6 & 40.6 \\
bol. luminosity & 42.0 & 43   &      & \\ 
Kinetic energy  & 57.5 & 55.7 & 59.8 & \\
Mass            & 9.5  & 7.3  &      & \\
$\tau_{diss}$    & 8.2  & 7.2  &      & \\
\hline
\end{tabular}
\caption{Overview of the luminosities of molecular line emission and
  H$\alpha$ (in erg s$^{-1}$), kinetic energies (in erg s$^{-1}$),
  masses (in M$_{\odot}$), and dissipation times (in yrs). We give the
  logarithms for all values. Column (1) -- Molecular gas measured with
  IRAM and Spitzer. The dissipation time is given under the assumption
  that all of the kinetic energy is dissipation through molecular line
  emission. Column (2) -- Warm ionized gas (the observed luminosity is
  that of H$\alpha$). The dissipation time is given under the
  assumption that all of the UV/optical line emission is due to the
  dissipation of kinetic energy. Column (3) -- Radio luminosity
  measured at 327 MHz. The kinetic energy was derived by Willis \&
  Strom (1978). Column (4) -- X-ray luminosity of the nuclear point
  source given by \citet{ogle09}.}
\label{tab:summary}
\end{table}

%%%%%%%%%%%%%%%%%%%%%%%%%%%%%%%%%%%%%%%%%%%%%%%%
%   TABLE: H2 lines FLUXES FOR 3C~326 predicted with shock models
%%%%%%%%%%%%%%%%%%%%%%%%%%%%%%%%%%%%%%%%%%%%%%%%
\renewcommand{\arraystretch}{1.1}
\begin{table}
\begin{center}
\begin{minipage}[t]{\textwidth}
\renewcommand{\footnoterule}{}
\centering
\small
\caption[MHD shock model parameters and predicted H$_2$ line fluxes for 3C~326]{MHD shock model parameters and predicted H$_2$ line fluxes$^{a}$ for 3C~326. }
   \begin{tabular}{c c c c c c c c c c c c}
   \hline
   \hline
n$_{\rm H}$$^b$  &  $V_s$ $^c$  &  \multicolumn{8}{c}{{\sc H$_2$ Line Fluxes} [$10^{-18}$ W~m$^{-2}$]} &  \multirow{2}*{$\mathcal{F}_{\rm H_2}$ $^d$} & \multirow{2}*{$\mathcal{F}_{\rm bol}$ $^e$} \\
\cline{3-10}
[cm$^{-3}$]  & $\rm [km~s^{-1}]$   & S(0) & S(1) & S(2) & S(3) & S(4) & S(5) & S(6) & S(7) & & \\
\hline
\multirow{4}*{$10^4$} & 4   &  0.74 & 4.51 & 0.18 & $6.1\! \times \!10^{-3}$ & $3\! \times \!10^{-4}$ & $8.6\! \times \!10^{-4}$ &$2.9\! \times \!10^{-4}$ &$7.5\! \times \!10^{-4}$ & 5.43 & 13.02    \\
& 15 & 0.09 &  3.22 &  3.19 &  10.36 &   1.35 &   0.68 &  0.02 & $2.2\! \times \!10^{-3}$ & 18.92& 20.55 \\
& 34 &  $4.2\! \times \!10^{-3}$ &  0.19 &  0.29 &   2.32 &   1.38 &   5.33 &   1.58 &  2.99 & 14.09 & 16.35  \\
\cline{2-12}
& Sum &  0.84 &  7.92 &   3.66 &  12.69 &  2.73 &  6.01 &  1.60 &   2.99 &  38.44 & 49.92   \\
\hline
\multirow{4}*{$10^3$} & 4 & 1.61 & 3.74 &  0.018 &  $1.3\! \times \!10^{-3}$ & $4.3\! \times \!10^{-4}$ & $1.2\! \times \!10^{-3}$ & $4\! \times \!10^{-4}$ & $10^{-3}$ & 5.37 & 10.6 \\
& 24 &  0.08 &  3.01 & 2.44 & 5.09 & 0.46 & 0.17 & $7\! \times \!10^{-3}$ & $6.6 \! \times \!10^{-3}$ & 11.3 & 11.6 \\
& 40 & 0.02 & 1.01 & 1.32 & 7.57 & 2.73 & 5.24 & 0.68 &  0.56 & 19.13 & 20.07 \\
\cline{2-12}
& Sum & 1.71 & 7.75 &  3.78 & 12.66 &   3.2 &  5.42 &  0.68 &  0.56 &  35.77 & 42.27\\
\hline
\multicolumn{2}{c}{\textit{Spitzer} Obs.} & $2.0\! \pm \!0.3$ & $7.5 \! \pm \! 0.5$ & $4.1 \! \pm \! 0.4$ & $12.6 \! \pm \! 0.4$ & $3.1 \! \pm \! 0.6$ & $5.0 \! \pm \!  2.5$ & $2.5 \! \pm \! 1.0$ & $2.9 \! \pm \! 1.0$ & $39.7 \! \pm \! 2.3$  &  \\
\hline
\end{tabular}
\label{tab:shock_H2_fluxes_3C326}
\footnotetext[1]{This table lists the shock model velocities, the predicted H$_2$ rotational line fluxes, and bolometric luminosities associated with each shock velocity components. Measured H$_2$ fluxes are taken from \citet{ogle07,ogle09}.}
\footnotetext[2]{Preshock hydrogen density.}
\footnotetext[3]{MHD Shock velocity.}
\footnotetext[4]{Sum of the $\rm H_2$ S(0) to S(7) rotational lines in $10^{-18}$ W~m$^{-2}$}
\footnotetext[5]{Sum over all the lines (bolometric luminosity of the shock)}
\normalsize
\end{minipage}
\end{center}
\end{table}
\renewcommand{\arraystretch}{1.0}

\begin{table}
\begin{center}
\renewcommand{\footnoterule}{}
\def\thefootnote{\alph{footnote}}
\centering
\begin{tabular}{lcccc}
\hline
$\rm n_H$ & $\rm V_s$ & Mass flow & t$_{\rm cool}$ & M$_{\rm gas}$ \\
(1)       & (2)       & (3)       & (4)           & (5) \\
\hline
$10^4$    & 4         & 70100     & 19050         & $1.3\times 10^9$\\
$10^4$    & 15        &  5950     & 3320          & $2.0\times 10^7$\\
$10^4$    & 34        & 730       & 1000          & $7.3\times 10^5$\\
\hline
$10^3$    & 4         & 41050     & 65000         & $2.7\times 10^9$\\
$10^3$    & 24        & 1100      & 11500         & $1.3\times 10^7$\\
$10^3$    & 40        & 630       &  5400         & $3.4\times 10^6$\\
\hline
\end{tabular}
\caption[1]{MHD shock model parameters, mass flows and cooling times. Column (1) -- Preshock density in cm$^{-3}$. Column (2) --
  Shock velocity in km s$^{-1}$. Column (3) -- Mass flow in M$_{\odot}$
  yr$^{-1}$. Column (4) -- Cooling time in yrs. Column (5) -- Gas mass in
  M$_{\odot}$. }
\label{tab:shock_mass_3C326}
\end{center}
\end{table}

\end{document}